\documentclass[oneside,english]{elsart}
\usepackage{ae,aecompl}
\usepackage[T1]{fontenc}
\usepackage[latin1]{inputenc}
\usepackage{subfigure}
\usepackage{amsmath}
\usepackage{graphicx}
\usepackage{amssymb}
\usepackage[authoryear]{natbib}

\makeatletter

\providecommand{\tabularnewline}{\\}

\usepackage{babel}
\makeatother

\begin{document}
\begin{frontmatter}

\title{The diffusive influx and carrier efflux have a strong effect on the
bistability of the \emph{lac} operon in \emph{Escherichia coli}}

\author{Jason T. Noel}

\address{Department of Chemical Engineering, University of Florida, Gainesville,
FL~32611-6005.}

\author{Sergei S. Pilyugin}

\address{Department of Mathematics, University of Florida, Gainesville, FL~32611-8105.}

\author{Atul Narang}

\address{Department of Chemical Engineering, University of Florida, Gainesville,
FL~32611-6005.}

\ead{narang@che.ufl.edu}

\thanks{Corresponding author. Tel: + 1-352-392-0028; fax: + 1-352-392-9513}

\begin{abstract}
In the presence of gratuitous inducers, the \emph{lac} operon of \emph{Escherichia
coli} exhibits bistability. Most models in the literature assume that
the inducer enters the cell via the carrier\emph{ }(permease), and
exits by a diffusion-like process. The diffusive influx and carrier
efflux are neglected. However, analysis of the data shows that in
non-induced cells, the diffusive influx is comparable to the carrier
influx, and in induced cells, the carrier efflux is 7 times the diffusive
efflux. Since bistability entails the coexistence of steady states
corresponding to both non-induced and induced cells, neither one of
these fluxes can be ignored. Here, we formulate a model accounting
for both fluxes. We show that: (a)~The thresholds of bistability
are profoundly affected by both fluxes. The diffusive influx reduces
the on threshold by enhancing inducer accumulation in non-induced
cells. The carrier efflux increases the off threshold by decreasing
inducer accumulation in induced cells. (b)~Simulations of the model
with experimentally measured parameter values are in good agreement
with the data for IPTG. However, there are discrepancies with respect
to the data for TMG. They are most likely due to two features missing
from the model, namely, the variation of the inducer exclusion effect
and the specific growth rate with the lactose enzyme levels. (c)~The
steady states and thresholds obtained in the presence of both fluxes
are well approximated by simple analytical expressions. These expressions
provide a rigorous framework for the preliminary design of the \emph{lac}
genetic switch in synthetic biology.
\end{abstract}
\end{frontmatter}

\section{Introduction}

The \emph{lac} operon is a paradigm of the mechanisms controlling
gene regulation. This interest was stimulated by the hope that insights
into the mechanism of \emph{lac} induction would shed light on the
central problem of development, namely, the mechanism by which genetically
identical cells acquire distinct phenotypes~\citep{monod47,Spiegelman1948}.

Many of the early studies of the \emph{lac} operon were concerned
with the kinetics of enzyme induction in the presence of \emph{gratuitous}
inducers, such as thiomethyl galactoside (TMG) and isopropyl thiogalactoside
(IPTG). The use of such inducers was important because they enabled
the kinetics of enzyme synthesis to be separated from the masking
effects of dilution. This was achieved by growing the cells in a medium
containing a gratuitous inducer (which promotes enzyme synthesis,
but not growth), and non-galactosidic carbon sources, such as succinate,
glycerol, and glucose (which support growth, but not enzyme synthesis).

\begin{figure}[t]
\noindent \begin{centering}
\subfigure[]{\includegraphics[width=2.5in]{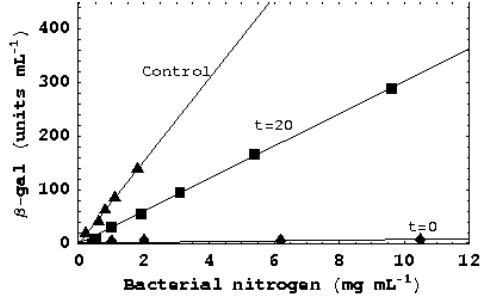}}
\par\end{centering}

\noindent \begin{centering}
\subfigure[]{\includegraphics[width=2.5in]{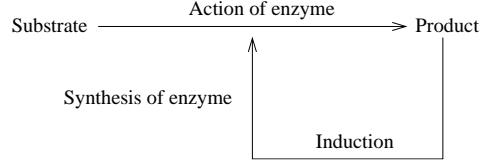}}
\par\end{centering}

\caption{\label{f:Cohn}(a)~Bistability during growth of \emph{E. coli} ML30
on succinate, glucose, and TMG~\citep[Fig.~4]{cohn59a}. If glucose
and TMG are added simultaneously to a culture growing on succinate,
there is no $\beta$-galactosidase synthesis ($\blacklozenge$). If
glucose is added 20~min after the addition of TMG, the enzyme is
synthesized ($\blacksquare$) at a rate that is 50\% of the rate observed
when only TMG is added to the culture ($\blacktriangle$). (b)~Kinetic
scheme proposed by Cohn and Horibata~\citep[p.~611]{cohn59a}. Here,
the enzyme represents \emph{lac} permease; the substrate and product
refer to extracellular and intracellular TMG, respectively.}

\end{figure}

These early studies showed that the enzyme synthesis rate was not
uniquely determined by the composition of the medium. If glucose and
TMG were added simultaneously to a culture of \emph{E. coli} ML30
growing on succinate, there was almost no synthesis of $\beta$-galactosidase;
however, if glucose was added 20~mins after the addition of TMG,
$\beta$-galactosidase was synthesized for up to 130 generations (Fig.~\ref{f:Cohn}a).
Thus, enzyme synthesis is bistable: Pre-induced cells remain induced,
and non-induced cells remain non-induced.

Cohn and Horibata found that the existence of bistability depended
crucially upon the existence of \emph{lac} permease (LacY): It disappeared
in \emph{lacY}$^{-}$ (cryptic), but not \emph{lac}Z$^{-}$, mutants~\citep{cohn59a}.
They proposed that bistability occurred due to the destabilizing effect
of the positive feedback generated by \emph{lac} permease. More precisely,
they observed that:

\begin{quotation}
This permease not only accumulates galactosides intracellularly, but
its synthesis is induced by the galactosides, which it accumulates.
The permease may therefore be described as a {}``self-inducing''
system, for which a generalized schema might be drawn.
\end{quotation}
This generalized schema, shown in Fig.~\ref{f:Cohn}b, was postulated
before the molecular mechanism of induction was discovered by Jacob
and Monod~\citep{jacob61}.

Some years after the discovery of the induction mechanism, Babloyantz
and Sanglier formulated a mathematical model, which was based on the
Cohn-Horibata scheme, but described enzyme induction in terms of the
now well established Jacob-Monod mechanism~\citep{Babloyantz1972}.
They showed that the model yielded the bistability observed in experiments.
Chung and Stephanopoulos formulated a similar model, the main difference
being that repressor-operator and repressor-inducer binding were assumed
to be in quasi-equilibrium~\citep{chung96}. This model is given
by the equations\begin{align}
\frac{dx}{dt} & =r_{s}-r_{x}-r_{g}x,\; r_{s}\equiv V_{s}e\frac{s}{K_{s}+s},\; r_{x}\equiv k_{x}\left(x-s\right).\label{eq:ChungX}\\
\frac{de}{dt} & =r_{e}^{+}-r_{e}^{-}-r_{g}e,\; r_{e}^{+}\equiv V_{e}\frac{1+K_{x}^{2}x^{2}}{1+\alpha+K_{x}^{2}x^{2}},\; r_{e}^{-}\equiv k_{e}^{-}e\label{eq:ChungE}\end{align}
where $x$ and $s$ denote the intracellular and extracellular TMG
concentrations, respectively; $e$ denotes the permease activity;
$r_{g}$ is the specific growth rate on the non-galactosidic carbon
source(s); $r_{s}$ is the specific rate of TMG uptake by the permease;
$r_{x}$ is the net specific rate of TMG efflux by diffusion; and
$r_{e}^{+},r_{e}^{-}$ denote the specific rates of permease synthesis
and degradation, respectively. The expression for $r_{e}^{+}$ is
based on a molecular model which assumes that the \emph{lac} operon
contains one operator, and the \emph{lac} repressor contains two identical
inducer-binding sites~\citep{yagil71}. The parameter, $K_{x}$,
is the association constant for the repressor-inducer binding; and
$\alpha$ is jointly proportional to the intracellular repressor level
and the association constant for repressor-operator binding. Evidently,
$\alpha$ is a measure of the \emph{repression}, defined as the ratio,
$\left.r_{e}^{+}\right|_{x\rightarrow\infty}/\left.r_{e}^{+}\right|_{x=0}$.

\begin{figure}[t]
\noindent \begin{centering}
\subfigure[]{\includegraphics[width=2.5in]{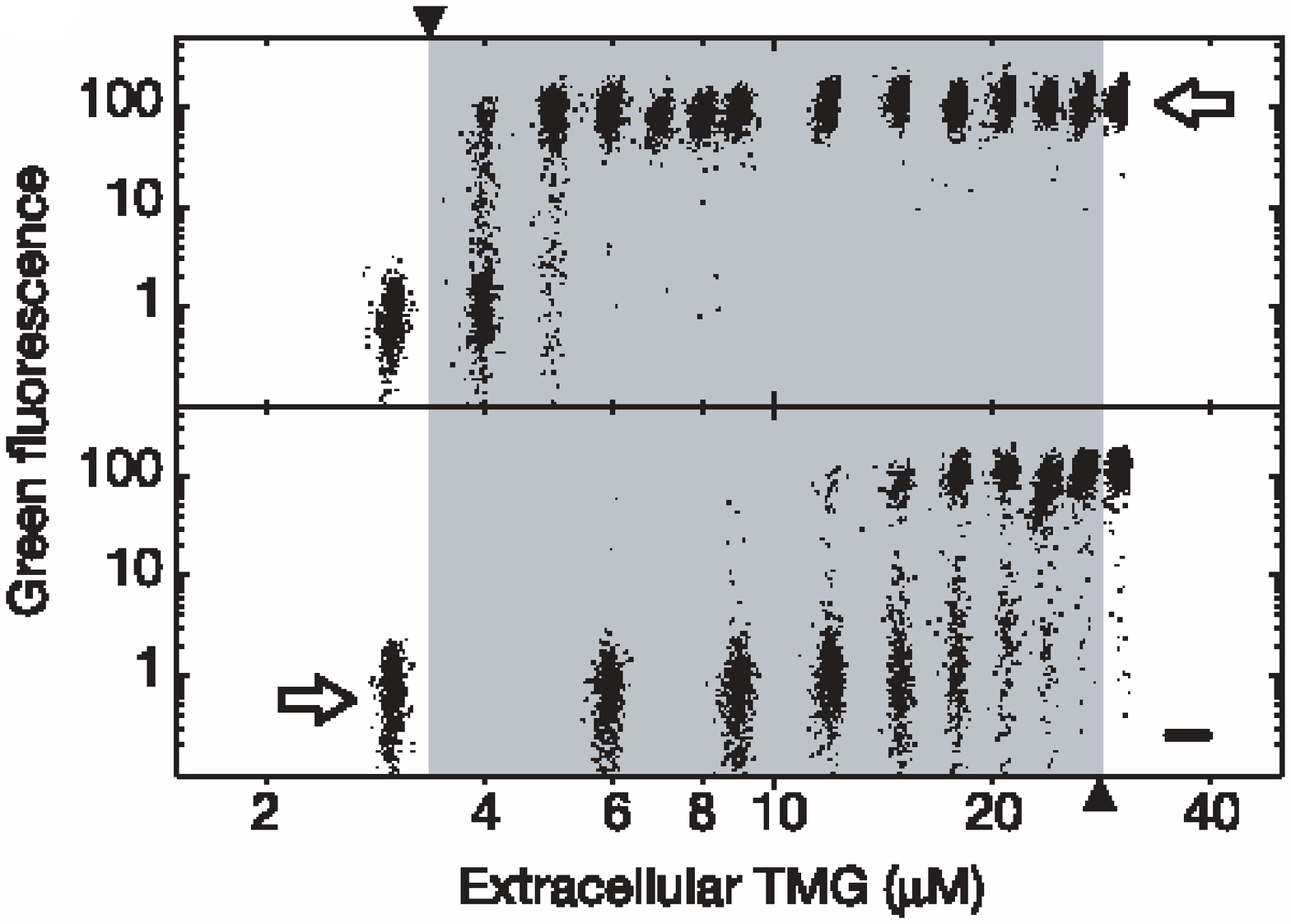}}\hspace*{0.2in}\subfigure[]{\includegraphics[width=2.5in]{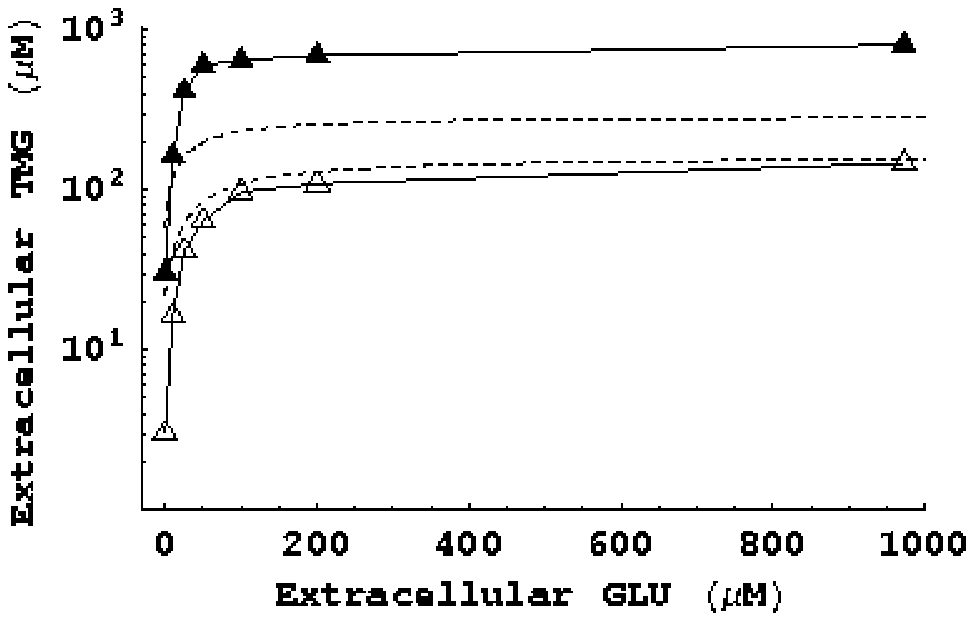}}
\par\end{centering}

\caption{\label{f:Oudenaarden1}Bistability of the \emph{lac} operon during
growth of \emph{E. coli} K12 MG1655 on succinate and succinate + glucose~\citep[Figs.~2b and~2c]{ozbudak04}.
(a)~Bistability during growth on succinate and various concentrations
of extracellular TMG. The (normalized) green fluorescence provides
a measure of the steady state activity of the \emph{lac} operon. The
upper (resp., lower) panel shows the green fluorescence observed when
an induced (resp., non-induced) inoculum is grown exponentially on
a mixture of succinate and various concentrations of extracellular
TMG. Bistability occurs at extracellular TMG concentrations between
the off and on thresholds of 3 and 30~$\mu$M, respectively, defined
as the extracellular TMG concentrations at which <5\% of the cells
are in their initial state. (b)~Bistability persists even if glucose
is added to the mixture of succinate and TMG. The lower ($\vartriangle$)
and upper ($\blacktriangle$) curves show the off and on thresholds
observed at various glucose concentrations. The dashed curves show
the thresholds predicted by the model with the parameter values in
Table~\ref{tab:Parameters}.}

\end{figure}

Although the experiments done by Cohn and coworkers provided clear
evidence of bistability during growth on TMG and succinate/glucose,
they did not investigate the variation of the enzyme levels with TMG
and glucose levels, and the effect of glucose on the key molecular
mechanisms of \emph{lac} regulation, namely, CRP-cAMP activation and
inducer exclusion. Recently, Ozbudak et al\@.~addressed these issues
by inserting two reporter operons into the chromosome of \emph{Escherichia
coli} K12 MG1655~\citep{ozbudak04}.

\begin{itemize}
\item The reporter \emph{lac} operon, placed under the control of the \emph{lac}
promoter, coded for green fluorescent protein (GFP) instead of the
\emph{lac} enzymes. Thus, the GFP intensity of a cell provided a measure
of the \emph{lac} enzyme level.
\item The reporter \emph{gat} operon, placed under the control of the constitutive
\emph{gat} promoter, coded for the red fluorescent protein (RFP) instead
of the \emph{gat} enzymes. The RFP intensity of a cell provided a
measure of its CRP-cAMP level.
\end{itemize}
To assess the effect of various TMG and glucose levels on \emph{lac}
bistability, they exposed non-induced and induced cells to a fixed
concentration of succinate, and various concentrations of TMG and
glucose. It was observed that:

\begin{enumerate}
\item When the cells were grown in the presence of succinate and various
concentrations of TMG, they exhibited bistability (Fig.~\ref{f:Oudenaarden1}a).
\item This bistability persisted even if glucose was added to the mixture
of succinate and TMG, but the thresholds increased with the concentration
of extracellular glucose (Fig.~\ref{f:Oudenaarden1}b).
\end{enumerate}
They also showed that both observations were mirrored by the bifurcation
diagram for the modified Chung-Stephanopoulos model, obtained by neglecting
the diffusive influx (i.e., the term, $k_{x}s$, in Eq.~\eqref{eq:ChungX}
was assumed to be zero).

\begin{figure}[t]
\noindent \begin{centering}
\subfigure[]{\includegraphics[width=2.5in]{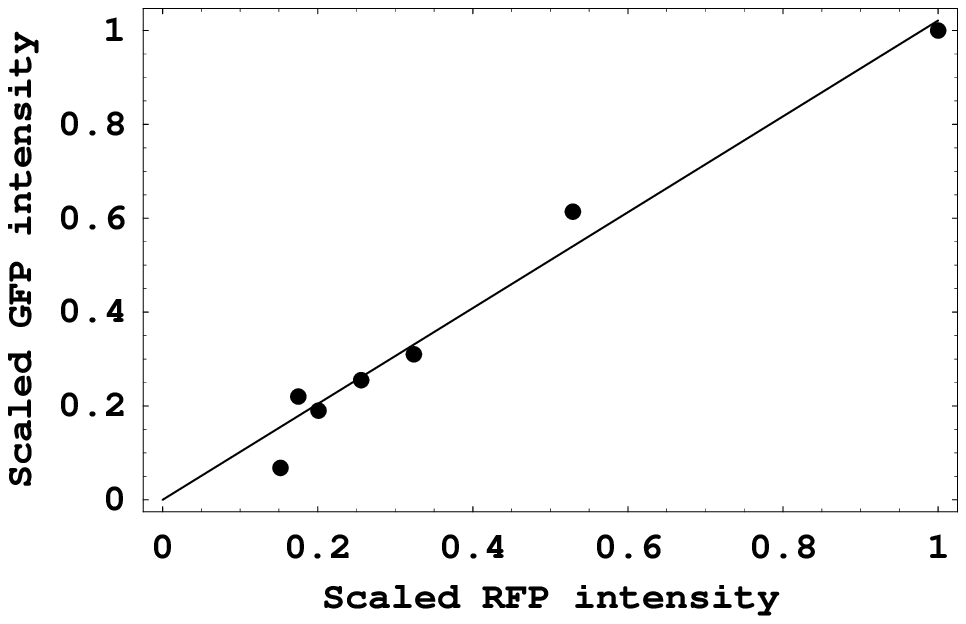}}\hspace*{0.2in}\subfigure[]{\includegraphics[width=2.5in]{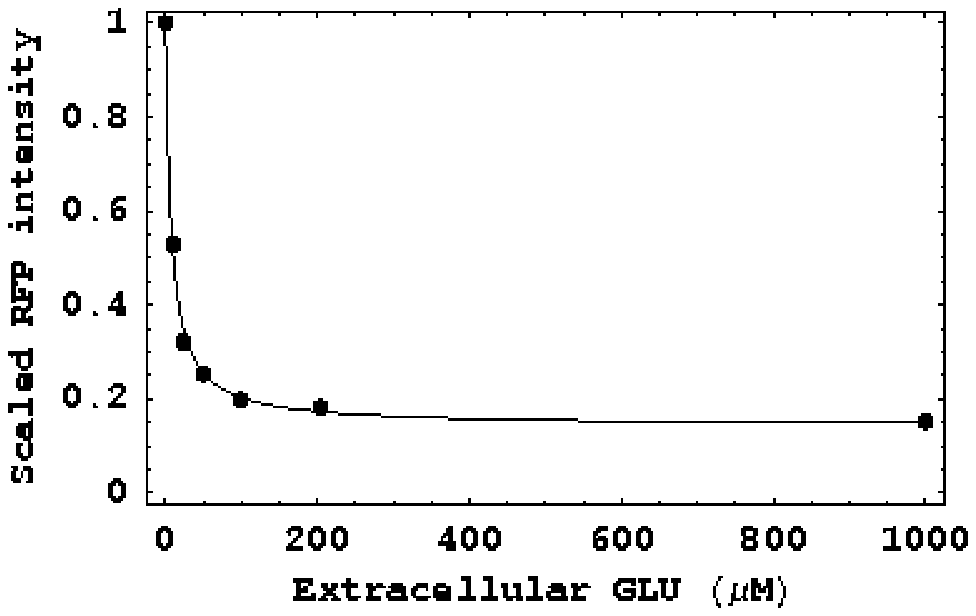}}
\par\end{centering}

\caption{\label{f:Oudenaarden2}Variation of the scaled red and green fluorescence
intensities during exponential growth of \emph{E. coli} K12 MG1655
in the presence of succinate, TMG, and various concentrations of glucose.
(a)~The scaled red and green fluorescence intensities are identical
at all glucose concentrations~\citep[Fig.~3b]{ozbudak04}. (b)~The
scaled red fluorescence intensity decreases with the concentration
of glucose~\citep[Fig.~3a]{ozbudak04}. The curve shows the fit to
Eq.~\eqref{eq:FitPhiEPhiG}.}

\end{figure}

To quantify the \emph{lac} repression due to glucose-mediated reduction
of CRP-cAMP levels, they measured the average green and red fluorescence
intensities of cells grown overnight in a medium containing a high
concentration of TMG, a fixed concentration of succinate, and various
concentrations of glucose. They found that the green and red fluorescence
intensities, scaled by the corresponding values observed in a medium
containing no glucose, were equal at all glucose concentrations (Fig.~\ref{f:Oudenaarden2}a).
Furthermore, the normalized RFP intensity decreased 5-fold at saturating
glucose concentrations (Fig.~\ref{f:Oudenaarden2}b). Ozbudak et
al.~assumed that the GFP intensity is proportional to the maximum
\emph{lac} induction rate, and the RFP intensity is proportional to
the CRP-cAMP level. Given these two assumptions, it was concluded
from Figs.~\ref{f:Oudenaarden2}a and~b that the maximum \emph{lac}
induction rate is proportional the CRP-cAMP level, and decreases 5-fold
at saturating concentrations of glucose. This is somewhat higher than
the maximum possible cAMP-mediated repression in \emph{E. coli} K12
MG1655. Recently, Kuhlman et al.~found that during growth of this
strain on saturating concentrations of 0.5\% glucose and 1~mM IPTG,
variation of the extracellular cAMP level from 0 to 10~mM produces
no more than a 3-fold change in \emph{lac} expression~\citep[Fig.~1B]{Kuhlman2007}.

The modified Chung-Stephanopoulos model considered by Ozbudak et al.~provides
an excellent point of departure for quantifying \emph{lac} bistability.
However, it suffers from three deficiencies.

\paragraph*{The induction kinetics are incorrect}

\begin{figure}[t]
\noindent \begin{centering}
\subfigure[]{\includegraphics[width=2.5in]{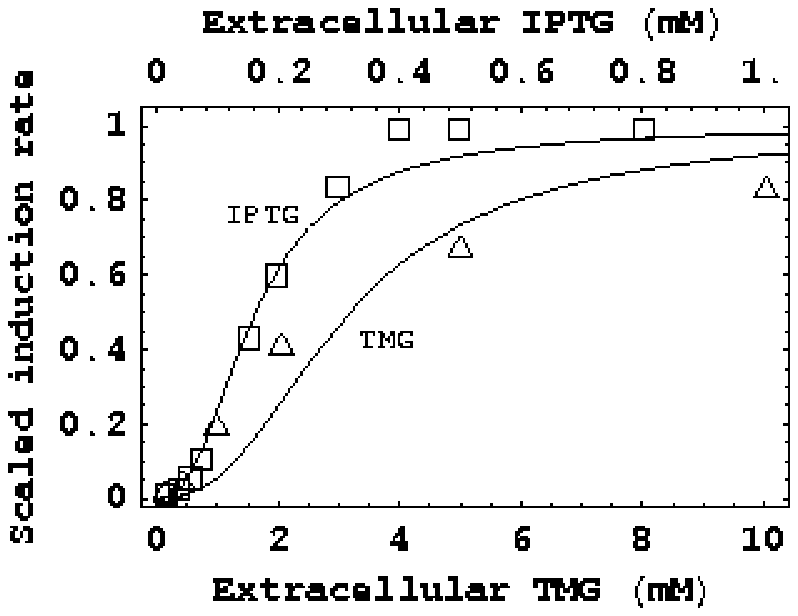}}\hspace{0.2in}\subfigure[]{\includegraphics[width=2.5in]{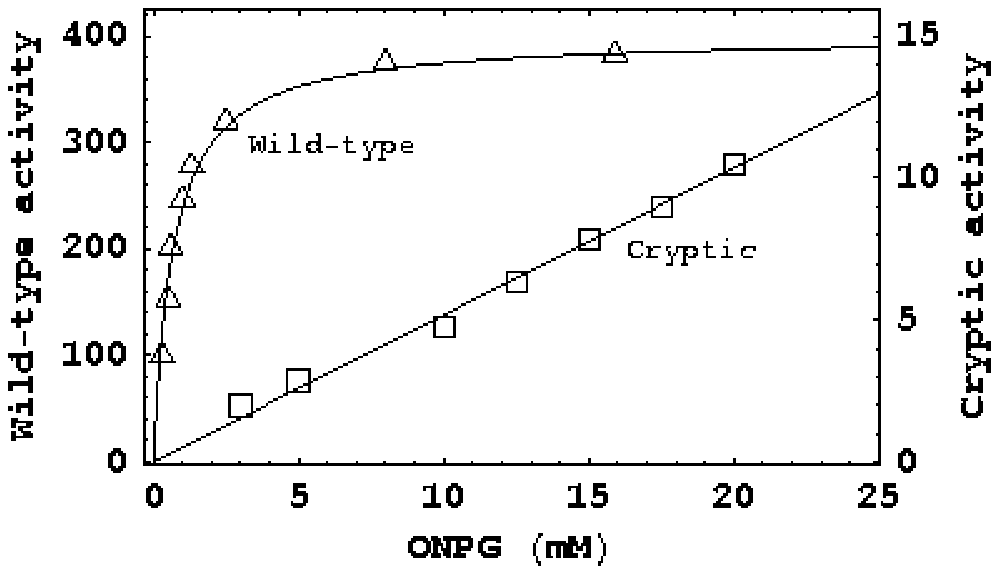}}
\par\end{centering}

\caption{\label{fig:Herzenberg}The kinetics of induction and diffusive influx
in the cryptic mutant, \emph{E. coli} ML3~\citep[Figs.~1 and 4]{Herzenberg1959}.
(a)~Variation of the scaled induction rate with extracellular TMG
and IPTG levels. The full curves show the fits to the data obtained
with Eq.~\eqref{eq:NewKinetics} and the parameter values $\alpha=40$,
$\hat{\alpha}=1200$, $K_{x}^{-1}=24$~$\mu$M for IPTG, and $K_{x}^{-1}=470$~$\mu$M
for TMG. (b)~Existence and kinetics of the diffusive influx. The
cryptic mutant hydrolyzes ONPG, and the hydrolysis rate increases
linearly with the extracellular TMG level. The permease activity is
expressed in $\mu$mols min$^{-1}$ gdw$^{-1}$.}

\end{figure}

The Yagil and Yagil model of \emph{lac} induction is not consistent
with the structure of the \emph{lac} operon and repressor. Indeed,
the \emph{lac} operon contains two auxiliary operators, $O_{2}$ and
$O_{3}$, in addition to the main operator, $O_{1}$, and the \emph{lac}
repressor contains four (not two) inducer-binding sites~\citep{Lewis2005}.
Furthermore, these structural features play a crucial role in the
formation of DNA loops, the key determinants of \emph{lac} repression~\citep{Oehler1990,Oehler1994}
and induction~\citep{Oehler2006}. Molecular models taking due account
of the 3 operators and 4 inducer-binding sites yield the \emph{lac}
induction rate\begin{equation}
r_{e}^{+}\equiv V_{e}\frac{1}{1+\alpha/\left(1+K_{x}x\right)^{2}+\hat{\alpha}/\left(1+K_{x}x\right)^{4}},\label{eq:NewKinetics}\end{equation}
where $K_{x}$ is the association constant for repressor-inducer binding,
and $\alpha,\hat{\alpha}$ are related to the the repression stemming
from repressor-operator binding and DNA looping, respectively~\citep{Kuhlman2007,Narang2007b}.
Equation~\eqref{eq:NewKinetics} yields good fits to the induction
curves obtained with various strains~\citep[Fig.~11]{Narang2007b}
and gratuitous inducers (Fig.~\ref{fig:Herzenberg}a).

\paragraph*{The diffusive influx is absent}

It is well known that there is a diffusive influx. Herzenberg showed
the existence of the diffusive influx by measuring the ONPG hydrolysis
rate in wild-type and cryptic cells.%
\footnote{The ONPG hydrolysis rate in intact cells is proportional to the rate
at which ONPG \citep{Herzenberg1959} or TMG \citep[Fig.~1]{Maloney1973}
permeate the cell.%
} He found that even cryptic cells hydrolyzed ONPG, and the hydrolysis
rate increased linearly with the extracellular ONPG concentration
(Fig.~\ref{fig:Herzenberg}b). It follows that gratuitous inducers
can enter the cells by a permease-independent mechanism, which can
formally be described by diffusive (first-order) kinetics.

Now, it may be permissible to neglect the diffusive influx in induced
cells because such cells contain such large permease levels that the
inducer is imported almost entirely by the permease. However, non-induced
cells contain only $\sim$0.1\% of the permease in induced cells~\citep{Maloney1973}.
In such cells, it seems unlikely that the diffusive influx is negligible.
As we show below, this conjecture is supported by the data.

\paragraph*{The carrier efflux is absent}

\begin{figure}[t]

\noindent \begin{centering}
\subfigure[]{\includegraphics[width=2.5in]{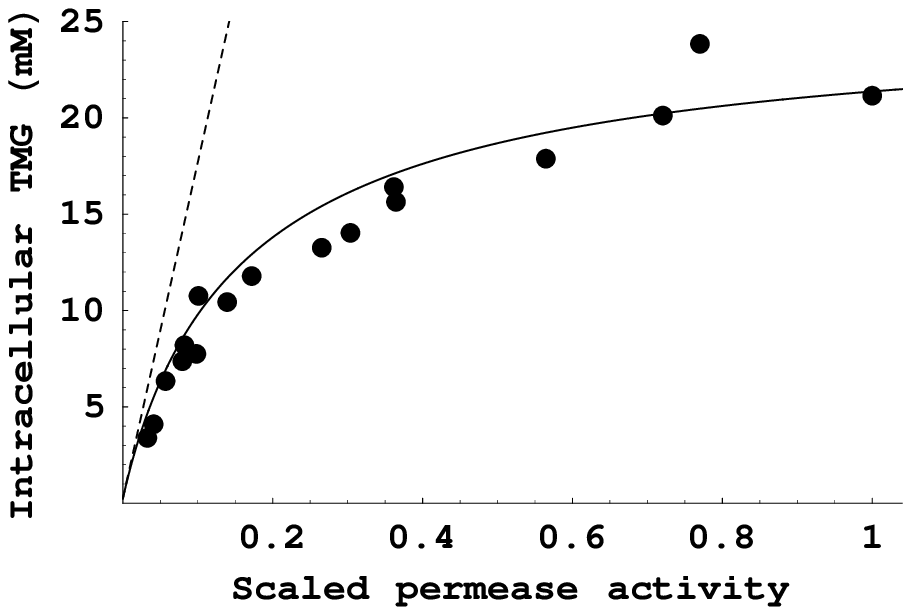}}\hspace{0.2in}\subfigure[]{\includegraphics[width=2.5in]{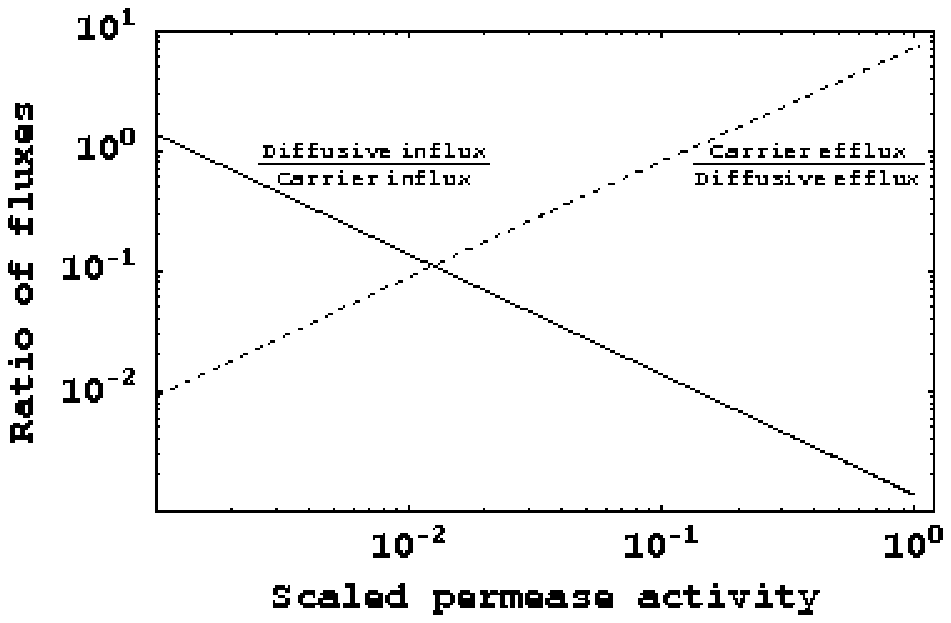}}
\par\end{centering}

\caption{\label{fig:Maloney}Existence and quantification of the carrier efflux.
(a)~When \emph{E. coli} K12 CA8000, induced to various levels, is
exposed to chloramphenicol and 0.24~M extracellular TMG, the steady
state intracellular TMG level increases nonlinearly with the permease
activity~\citep[Fig.~5]{Maloney1973}. The permease activity is normalized
by the activity of fully induced cells (43.5~$\mu$moles TMG per
min per mL of cell water). The full curve shows the fit obtained from
Eq.~\eqref{eq:MaloneyX} with the experimentally measured parameters,
$s=0.24$~mM, $K_{1}=0.8$~mM, $k_{x}=0.14$~min$^{-1}$, and the
best fit parameter estimate, $K_{2}=84$~mM. The dashed line shows
the intracellular TMG level predicted by Eq.~\eqref{eq:xLinearIncrease}.
(b)~Flux ratios estimated from~\eqref{eq:MaloneyX}. In fully induced
cells, the carrier efflux is $\sim7$ times the diffusive efflux.
In non-induced cells, which have a scaled permease activity of $\sim$0.001,
the diffusive influx is comparable to the carrier influx. }

\end{figure}

The existence of the carrier efflux was suggested in early studies
with gratuitous inducers~\citep{Koch1964}. However, Maloney and
Wilson were the first to quantify its effect~\citep{Maloney1973}.
To this end, they measured the steady state intracellular TMG levels
in cells induced to various levels, and then exposed to chloramphenicol
plus 0.24~mM extracellular TMG. The data obtained, shown in Fig.~\ref{fig:Maloney}a,
cannot be reconciled with models, such as the Chung-Stephanopoulos
model, that account only for the carrier influx and diffusion. Indeed,
since enzyme synthesis is blocked in the presence of chloramphenicol,
the Chung-Stephanopoulos model implies that the steady state intracellular
TMG level is given by the expression\begin{equation}
x\approx s+\left(\frac{V}{k_{x}}\right)\frac{s}{K_{1}+s},\; V\equiv V_{s}e,\label{eq:xLinearIncrease}\end{equation}
obtained from \eqref{eq:ChungX} by ignoring the negligibly small
dilution term. It follows that if cells containing various permease
levels are exposed to chloramphenicol and a fixed extracellular TMG
level, the intracellular TMG level, $x$, should increase linearly
with $V$, the specific activity of the permease. However, the experiments
show that $x$ increases linearly only if $V$ is small --- it saturates
at large values of $V$ (closed circles in Fig.~\ref{fig:Maloney}a).

To resolve this discrepancy, Maloney and Wilson postulated that the
permease catalyzes both influx and efflux, i.e., the steady state
intracellular TMG level satisfies the equation\begin{equation}
0=V\left(\frac{s}{K_{1}+s}-\frac{x}{K_{2}+x}\right)-k_{x}\left(x-s\right),\label{eq:MaloneyX}\end{equation}
where the term, $Vx/(K_{2}+x)$, accounts for the carrier efflux.
They measured $K_{1}$, $k_{x}$, and showed that \eqref{eq:MaloneyX}
yields a good fit to the data, provided $K_{2}$ is chosen to be 84~mM
(full curve in Fig.~\ref{fig:Maloney}a).

With the help of Eq.~\eqref{eq:MaloneyX}, they explained their data
as follows. If the cells have low enzyme levels, the intracellular
TMG levels are so small compared to $K_{2}$ that the carrier efflux
is negligible. Thus, \eqref{eq:MaloneyX} reduces to \eqref{eq:xLinearIncrease},
and the intracellular level increases linearly with $V$ (dashed line
in Fig.~\ref{fig:Maloney}a). However, at large enzyme levels, the
intracellular TMG level becomes so high that the steady state is completely
determined by the first two terms of \eqref{eq:MaloneyX}, namely,
the carrier influx and efflux. The intracellular TMG level is therefore
independent of the enzyme activity.

The foregoing physical argument assumes that the carrier efflux is
negligible in non-induced cells, and diffusion is negligible in induced
cells. These assumptions are consistent with the mathematical model.
Indeed, given Eq.~\eqref{eq:MaloneyX} and the above parameter values,
we can plot the two ratios, carrier efflux:diffusive efflux and diffusive
influx:carrier influx, as a function of the enzyme activity (Fig.~\ref{fig:Maloney}b).
The plots show that in induced cells, the carrier influx is $\sim$1000
times the diffusive influx (full line), and the carrier efflux is
almost $\sim$7 times the diffusive efflux (dashed line). In non-induced
cells, on the other hand, the carrier efflux is negligible compared
to the diffusive efflux, but the diffusive influx is comparable to
the carrier influx.

It is therefore clear that the carrier efflux cannot be ignored in
induced cells, and the diffusive influx cannot be neglected in non-induced
cells. Since bistability entails the coexistence of steady states
corresponding to both induced and non-induced cells, neither the carrier
efflux nor the diffusive influx can be neglected. Yet, one or both
the fluxes have been ignored in most models of \emph{lac} bistability~\citep{chung96,Hoek2006,Narang2007c,ozbudak04,Santillan2007,Tian2005}.
To be sure, both fluxes were included in one of the models~\citep{Vilar2003a},
but their influence on bistability was not analyzed. The goal of this
work is analyze this effect.

Recently, we studied the effect of DNA looping on the dynamics of
the modified Chung-Stephanoupoulos model~\citep{Narang2007c}. Here,
we formulate and analyze an extended model of \emph{lac} transcription
that takes due account of DNA looping as well as the diffusive influx
and the carrier efflux. We find that:

\begin{enumerate}
\item The diffusive influx has no effect on the off threshold, but it significantly
reduces the on threshold.
\item The carrier efflux has no effect on the on threshold, but it significantly
increases the off threshold.
\end{enumerate}
We show that the on and off thresholds can be represented by simple
analytical expressions.

\section{Model}

\begin{figure}[t]
\noindent \begin{centering}
\includegraphics[width=4in]{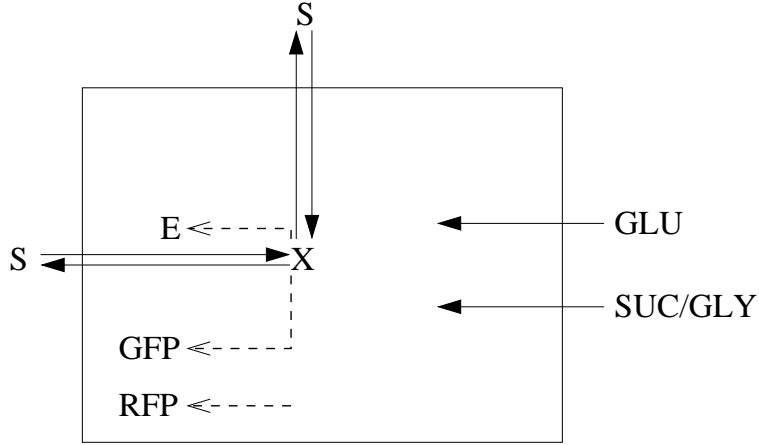}
\par\end{centering}

\caption{\label{f:Scheme}Kinetic scheme of the model. }

\end{figure}

Fig.~\ref{f:Scheme} shows the kinetic scheme of the model. Here,
$E$ denotes \emph{lac} permease; $S$ and $X$ denote the extracellular
and intracellular inducer, respectively; GFP and RFP denote the green
and red fluorescent protein synthesized by the reporter \emph{lac}
and constitutive \emph{gat} operons, respectively; and GLU, SUC, GLY
denote glucose, succinate, and glycerol. We assume that:

\begin{enumerate}
\item The concentrations of the permease, intracellular inducer, GFP, and
RFP, denoted $e$, $x$, $g$, and $r$, respectively, are based on
the volume of the cells (mol~L$^{-1}$). The concentrations of the
extracellular inducer and glucose, denoted $s$ and $G$, respectively,
are based on the volume of the reactor (mol~L$^{-1}$). The rates
of all the processes are based on the volume of the cells (mol~h$^{-1}$~~L$^{-1}$).
We shall use the term \emph{specific }to emphasize this point.\\
The choice of these units implies that if the concentration of any
intracellular component, $Z$, is $z$ mol~L$^{-1}$, then the evolution
of $z$ in batch cultures is given by\[
\frac{dz}{dt}=r_{z}^{+}-r_{z}^{-}-r_{g}z\]
where $r_{z}^{+}$ and $r_{z}^{-}$ denote the specific rates of synthesis
and degradation of $Z$ in mol~h$^{-1}$~L$^{-1}$, and $r_{g}$
is the specific growth rate in h$^{-1}$.
\item The specific growth rate is given by\[
r_{g}\equiv r_{g0}\phi_{g}(G),\]
where $r_{g0}$ denotes the specific growth rate in the absence of
glucose (i.e., in the presence of pure succinate), and $\phi_{g}(G)$
is an increasing function of the extracellular glucose concentration,
$G$, satisfying $\phi_{g}(0)=1$, which accounts for the increase
of the specific growth rate produced by addition of glucose to a culture
growing on succinate.
\item The saturation constant for carrier efflux, $K_{2}$, is so large
that $x/(K_{2}+x)\approx x/K_{2}$. Hence, the net carrier-mediated
specific uptake rate of $S$ is \[
r_{s}\equiv V_{s0}\phi_{s}(G)e\left(\frac{s}{K_{1}+s}-\frac{x}{K_{2}}\right),\]
where $V_{s0}e$ is the specific activity of the permease in the absence
of glucose, and $\phi_{s}(G)$ is a decreasing function of $G$ such
that $\phi_{s}(0)=1$, which accounts for the reduction of the specific
activity due to inducer exclusion.
\item The net specific rate of expulsion of $X$ by diffusion follows the
kinetics \[
r_{x}\equiv k_{x}\left(x-s\right).\]

\item The intracellular inducer stimulates the transcription of the native
and reporter \emph{lac} operons, resulting in the synthesis of the
\emph{lac} enzymes and GFP, respectively.

\begin{enumerate}
\item The specific synthesis rate of the permease, $E$, is\begin{equation}
r_{e}\equiv V_{e0}\phi_{e,lac}(G)\frac{1}{1+\alpha/\left(1+K_{x}x\right)^{2}+\hat{\alpha}/\left(1+K_{x}x\right)^{4}},\label{eq:Induction}\end{equation}
where $V_{e0}$ is the specific synthesis rate in fully induced cells
growing on succinate ($G=0$); $\phi_{e,lac}(G)$ is a decreasing
function of $G$ satisfying $\phi_{e,lac}(0)=1$, which accounts for
the repression of transcription due to reduction of CRP-cAMP levels
in the presence of glucose; $K_{x}$ is the association constant for
repressor-inducer binding; and $\alpha,\hat{\alpha}$ characterize
the repression due to repressor-operator binding and DNA looping,
respectively.\\
The values of $\alpha$ and $\hat{\alpha}$ are 20--50 and $\sim$1200,
respectively, which reflect the fact that more than 95\% of the total
repression, $1+\alpha+\hat{\alpha}$, is due to the formation of DNA
loops~\citep{Oehler1990,Oehler1994}. The dissociation constant for
repressor-inducer binding, $K_{x}^{-1}$, is 7--30~$\mu$M for IPTG~\citep[Table~1]{Narang2007b}.
It is 10 times higher for TMG, based on a simple analysis of the data
with Scatchard plots~\citep[Table~I]{Barkley1975}. A similar estimate
is obtained when Eq.~\eqref{eq:NewKinetics} is used to fit the data
shown in Fig.~\ref{fig:Herzenberg}a. Assuming $\alpha=40$ and $\hat{\alpha}=1200$,
the best fits to this data are obtained when the dissociation constants
for IPTG and TMG are assumed to be 24 and 470~$\mu$M, respectively.
\item The specific synthesis rate of GFP is\[
r_{GFP}\equiv V_{e0}\phi_{e,lac}(G)\frac{1}{1+\alpha/\left(1+K_{x}x\right)^{2}+\hat{\alpha}_{G}/\left(1+K_{x}x\right)^{4}},\]
where $\hat{\alpha}_{G}<\hat{\alpha}$. This expression is obtained
by assuming that the promoters of the reporter and native \emph{lac}
operons are identical, so that both operons have the same maximum
specific synthesis rates in the absence of glucose, and are subject
to the very regulation by CRP-cAMP and repressor-operator binding.
However, the two operons differ with respect to regulation by DNA
looping because the reporter \emph{lac} operon lacks the auxiliary
operator, $O_{2}$. This precludes the formation of DNA loops by interaction
between $O_{1}$ and $O_{2}$.\\
Ozbudak et al.~estimated the total repression of the \emph{lac} reporter,
$1+\alpha+\hat{\alpha}_{G}$, to be 170 \citep{ozbudak04}, which
implies that $\hat{\alpha}_{G}\approx130$.
\end{enumerate}
\item The specific synthesis rate of RFP follows the constitutive kinetics\[
r_{RFP}\equiv V_{r0}\phi_{e,gat}(G),\]
where $V_{r0}$ is the specific synthesis rate of RFP in the absence
of glucose, and $\phi_{e,gat}(G)$ is a decreasing function of $G$
such that $\phi_{e,gat}(0)=1$, which accounts for the CRP-cAMP effect
exerted in the presence of glucose.
\item Protein degradation is negligible.
\end{enumerate}
Given these assumptions, the mass balances yield\begin{align}
\frac{dx}{dt} & =V_{s0}\phi_{s}e\left(\frac{s}{K_{1}+s}-\frac{x}{K_{2}}\right)-k_{x}\left(x-s\right)-r_{g0}\phi_{g}x,\label{eq:OrigX}\\
\frac{de}{dt} & =V_{e0}\phi_{e,lac}\frac{1}{1+\alpha/\left(1+K_{x}x\right)^{2}+\hat{\alpha}/\left(1+K_{x}x\right)^{4}}-r_{g0}\phi_{g}e,\label{eq:OrigE}\\
\frac{dg}{dt} & =V_{e0}\phi_{e,lac}\frac{1}{1+\alpha/\left(1+K_{x}x\right)^{2}+\hat{\alpha}_{G}/\left(1+K_{x}x\right)^{4}}-r_{g0}\phi_{g}g,\label{eq:OrigG}\\
\frac{dr}{dt} & =V_{r0}\phi_{e,gat}-r_{g0}\phi_{g}r.\label{eq:OrigR}\end{align}
It is convenient to rescale these equations. Ozbudak et al.~scaled
the GFP and RFP intensities by the corresponding intensities observed
during steady exponential growth on succinate and excess TMG, but
no glucose. Under these conditions, Eqs.~\eqref{eq:OrigE}--\eqref{eq:OrigR}
imply that $e$, $g$, and $r$ have the values $V_{e0}/r_{g0}$,
$V_{e0}/r_{g0}$, and $V_{r0}/r_{g0}$, respectively. Thus, we are
led to define the dimensionless variables\[
\epsilon\equiv\frac{e}{V_{e0}/r_{g0}},\;\gamma\equiv\frac{g}{V_{e0}/r_{g0}},\rho\equiv\frac{r}{V_{r0}/r_{g0}}.\]
If we scale the remaining variables, $x$ and $t$, as follows,\[
\chi\equiv\frac{x}{K_{x}^{-1}},\;\tau\equiv\frac{t}{r_{g0}^{-1}},\]
we arrive at the dimensionless equations\begin{align}
\tau_{x}\frac{d\chi}{d\tau} & =\delta_{m0}\phi_{s}\epsilon\left(\frac{\sigma}{\kappa_{1}+\sigma}-\frac{\chi}{\kappa_{2}}\right)-\left(\chi-\sigma\right)-\tau_{x}\phi_{g}\chi\label{eq:x}\\
\frac{d\epsilon}{d\tau} & =\phi_{e,lac}f(\chi)-\phi_{g}\epsilon,\; f(\chi)\equiv\frac{1}{1+\alpha/\left(1+\chi\right)^{2}+\hat{\alpha}/\left(1+\chi\right)^{4}},\label{eq:e}\\
\frac{d\gamma}{d\tau} & =\phi_{e,lac}h(\chi)-\phi_{g}\gamma,\; h(\chi)\equiv\frac{1}{1+\alpha/\left(1+\chi\right)^{2}+\hat{\alpha}_{G}/\left(1+\chi\right)^{4}},\label{eq:g}\\
\frac{d\rho}{d\tau} & =\phi_{e,gat}-\phi_{g}\rho,\label{eq:r}\end{align}
with the dimensionless parameters, \begin{equation}
\tau_{x}\equiv\frac{r_{g0}}{k_{x}},\;\sigma\equiv\frac{s}{K_{x}^{-1}},\;\kappa_{i}\equiv\frac{K_{i}}{K_{x}^{-1}},\;\delta_{m0}\equiv\frac{V_{s0}\left(V_{e0}/r_{g0}\right)/k_{x}}{K_{x}^{-1}}.\label{eq:ScaledParam}\end{equation}
Here, $\tau_{x}$ is the ratio of the time constant for inducer expulsion
by diffusion ($k_{x}^{-1}$) relative to the time constant for growth
on succinate ($r_{g0}^{-1}$); $\sigma$ is the extracellular inducer
concentration, measured in units of $K_{x}^{-1}$; $\kappa_{1}$ and
$\kappa_{2}$ are the saturation constants for carrier uptake and
expulsion, respectively, measured in units of $K_{x}^{-1}$.

The parameter, $\delta_{m0}$, is related to the  steady state intracellular
inducer level attained in fully induced cells exposed to chloramphenicol,
succinate, and saturating inducer levels, if\emph{ }the carrier efflux
is somehow abolished. Indeed, since $r_{g,0}=0.4$~h$^{-1}$, $k_{x}=0.14$~min$^{-1}$,
and $\phi_{g}\lesssim2$, the dilution term, $\tau_{x}\phi_{g}\chi$,
in \eqref{eq:x} is negligibly small compared to the diffusive efflux,
$\chi$. The steady states are therefore given by the relations\begin{align}
\chi & =\frac{\sigma+\delta_{m0}\phi_{s}\epsilon\sigma/(\kappa_{1}+\sigma)}{1+\delta_{m0}\phi_{s}\epsilon/\kappa_{2}},\label{eq:xSS}\\
\epsilon & =\frac{\phi_{e,lac}}{\phi_{g}}f(\chi),\label{eq:eSS}\\
\gamma & =\frac{\phi_{e,lac}}{\phi_{g}}h(\chi),\label{eq:gSS}\\
\rho & =\frac{\phi_{e,gat}}{\phi_{g}}.\label{eq:rSS}\end{align}
Equation \eqref{eq:xSS} provides the steady state intracellular inducer
level in cells exposed to fixed concentrations of chloramphenicol
and succinate, and any given concentrations of extracellular inducer
and glucose. If the carrier efflux is somehow abolished ($\kappa_{2}\rightarrow\infty$),
and the medium contains no glucose ($\phi_{s}=1$), the steady state
intracellular inducer level is given by the equation \begin{equation}
\chi=\sigma+\delta_{m0}\epsilon\frac{\sigma}{\kappa_{1}+\sigma}.\label{eq:ChiDeltaM}\end{equation}
If the cells are fully induced ($\epsilon=1$) and the extracellular
inducer level is saturating ($\sigma\gg\kappa_{1}$), the above equation
becomes\[
\chi=\sigma+\delta_{m0}\approx\delta_{m0},\]
where the first term was neglected because in induced cells, the diffusive
influx is vanishingly small compared to the carrier influx. Thus,
$\delta_{m0}$ is the steady state intracellular inducer level, measured
in units of $K_{x}^{-1}$, that is attained in fully induced cells
exposed to chloramphenicol, succinate, and saturating levels of extracellular
inducer,\emph{ }provided the carrier efflux is somehow abolished.

We can estimate the value of $\delta_{m0}K_{x}^{-1}$ by observing
that \eqref{eq:ChiDeltaM} is the dimensionless analog of \eqref{eq:xLinearIncrease}.
Indeed, multiplying \eqref{eq:ChiDeltaM} by $K_{x}^{-1}$ yields
the relation\[
x=s+K_{x}^{-1}\delta_{m0}\epsilon\frac{s}{K_{1}+s},\]
which, upon comparison with \eqref{eq:xLinearIncrease}, gives\[
K_{x}^{-1}\delta_{m0}\epsilon=\frac{V}{k_{x}}.\]
Since $V=43.5$~$\mu$moles min$^{-1}$ mL$^{-1}$ in fully induced
cells, we conclude that $\delta_{m0}K_{x}^{-1}=V/k_{x}=310$~mM.

If the medium contains the inducer, succinate, and possibly glucose,
but no chloramphenicol, enzyme synthesis is not blocked. Instead,
it responds to the prevailing intracellular inducer concentration,
and the enzyme level evolves toward the steady state given by \eqref{eq:eSS}.
In this case, the steady state intracellular inducer level satisfies
the relation\begin{equation}
\chi=\frac{\sigma+\delta_{m}(G)f(\chi)\sigma/(\kappa_{1}+\sigma)}{1+\delta_{m}(G)f(\chi)/\kappa_{2}},\delta_{m}(G)\equiv\delta_{m0}\phi(G),\;\phi(G)\equiv\frac{\phi_{s}\phi_{e,lac}}{\phi_{g}},\label{eq:EqCondn}\end{equation}
obtained by substituting \eqref{eq:eSS} in \eqref{eq:xSS}. Here,
$\phi(G)\le1$ captures the cumulative effect of glucose due to inducer
exclusion, catabolite repression, and enzyme dilution; and $\delta_{m}(G)$
is a measure of the steady state intracellular inducer level in fully
induced cells exposed to succinate, glucose, and saturating inducer
levels.

We are particularly interested in:

\begin{enumerate}
\item The variation of the steady states with the extracellular inducer
concentration at any given glucose level (for example, $G=0$, in
Fig.~\ref{f:Oudenaarden1}a). This is completely determined by Eq.~\eqref{eq:EqCondn}:
Given any $\sigma$ and $G$, the equation can be solved for $\chi$,
which can then be substituted in \eqref{eq:eSS}--\eqref{eq:gSS}
to obtain the corresponding $\epsilon$ and $\gamma$. We shall refer
to \eqref{eq:EqCondn} as the equilibrium\emph{ }condition.
\item The variation of the thresholds with the extracellular glucose concentration
(Fig.~\ref{f:Oudenaarden1}b). These thresholds are the points at
which the steady state bifurcates (i.e., changes its stability or
multiplicity). It is shown in Appendix~\ref{app:BifnCondn} that
a steady state bifurcates only if it satisfies the relation \begin{equation}
1+\delta_{m}\left[\frac{1}{\kappa_{2}}\left(f+\chi f_{\chi}\right)-\frac{\sigma}{\kappa_{1}+\sigma}f_{\chi}\right]=0.\label{eq:BifnCondn}\end{equation}
We shall refer to this equation as the bifurcation\emph{ }condition.
\end{enumerate}
Thus, the steady states are completely determined by the equilibrium
condition, whereas the thresholds must satisfy the equilibrium and
bifurcation conditions.

\section{Results and Discussion}

\subsection{The effect of dilution}

To simulate the effect of the diffusion influx and the carrier efflux,
we need the function, $\phi(G)\equiv\phi_{s}\phi_{e}/\phi_{g}$. In
the course of estimating this function, we shall also resolve the
discrepancy between the magnitudes of the cAMP-mediated repression
reported by Ozbudak et al.~and Kuhlman et al.

It follows from \eqref{eq:gSS} and \eqref{eq:rSS} that when the
cells are grown in the presence of excess TMG, the ratio of the steady
state GFP and RFP intensities is given by\[
\frac{\gamma}{\rho}=\frac{\phi_{e,lac}(G)}{\phi_{e,gat}(G)}.\]
Since this ratio was observed to be 1 all glucose concentrations (Fig.~\ref{f:Oudenaarden2}a),
the \emph{lac} and \emph{gat} operons respond identically to CRP-cAMP,
i.e., \begin{equation}
\phi_{e,lac}=\phi_{e,gat}=\phi_{e},\textnormal{say.}\label{eq:phiGatphiLac}\end{equation}
Moreover, the steady state RFP and GFP intensities are given by the
equation \begin{equation}
\rho=\gamma=\frac{\phi_{e}}{\phi_{g}}.\label{eq:CorrectlRho}\end{equation}
It follows that the 5-fold decline of $\rho$ and $\gamma$ at saturating
glucose concentrations, shown in Fig.~\ref{f:Oudenaarden2}b, represents
the combined effect of reduced CRP-cAMP levels and enhanced dilution
(as opposed to the the sole effect of reduced CRP-cAMP levels).

The effect of the reduced CRP-cAMP level, $\phi_{e}$, cannot be estimated
unless the effect of enhanced dilution, $\phi_{g}$, is known. Ozbudak
et al.~did not report the specific growth rates at various glucose
concentrations, but experiments show that \citep{narang97a}:

\begin{itemize}
\item When \emph{E. coli} K12 is grown on saturating levels of succinate
+ glucose, the cells consume only glucose during the first exponential
growth phase.
\item The maximum specific growth rate on glucose (0.74~h$^{-1}$) is roughly
2 times that on succinate (0.44~h$^{-1}$).
\end{itemize}
It follows that at saturating glucose concentrations, $\phi_{g}\approx2$.
Since $\rho\approx1/5$ under these conditions, \eqref{eq:CorrectlRho}
yields $\phi_{e}=\rho\phi_{g}=2/5$. Thus, roughly half of the 5-fold
decline in Fig.~\ref{f:Oudenaarden2}b is due to the enhanced dilution
rate at saturating glucose concentrations. Reduction of the CRP-cAMP
levels accounts for the remaining 2.5-fold decline, which is consistent
with the data obtained by Kuhlman et al.

In view of \eqref{eq:CorrectlRho}, we can assume that\begin{equation}
\frac{\phi_{e}}{\phi_{g}}=1-0.84\frac{G}{7.6+G},\label{eq:FitPhiEPhiG}\end{equation}
which represents the best fit obtained to the data in Fig.~\ref{f:Oudenaarden2}b.

It remains to specify the function, $\phi_{s}$, which characterizes
the intensity of inducer exclusion. We assume that the saturation
constants for inducer exclusion and cAMP activation/dilution are the
same (7.6~$\mu$M). Experiments show that the lactose uptake rate
decreases 2-fold in the presence of high glucose concentrations~\citep[Fig.~3]{McGinnis1969}.
Thus, we are led to postulate the expression,\begin{equation}
\phi_{s}=1-0.5\frac{G}{7.6+G}.\label{eq:FitPhiS}\end{equation}
We shall discuss the implications of the foregoing assumption later
on.

\begin{table}[t]
\noindent \begin{centering}
\caption{\label{tab:Parameters}Parameter values used in the simulations. All
the parameter values, except $K_{x}^{-1}$, are assumed to be the
same for TMG and IPTG. }

\par\end{centering}

\medskip{}

\noindent \begin{centering}
\begin{tabular}{|c|c|}
\hline
Parameter & Value and reference\tabularnewline
\hline
\hline
$K_{1}$ & 0.68 mM \citep{ozbudak04}\tabularnewline
\hline
$K_{2}$ & 84 mM \citep{Maloney1973}\tabularnewline
\hline
$K_{x}^{-1}$ & 0.3 mM for TMG \citep{Herzenberg1959}\tabularnewline
\hline
 & 0.007 mM for IPTG \citep{Oehler2006}\tabularnewline
\hline
$\alpha$ & 40 \citep{Oehler1994}\tabularnewline
\hline
$\hat{\alpha}$ & 1200 \citep{Oehler1994}\tabularnewline
\hline
$\hat{\alpha}_{G}$ & 130 \citep{ozbudak04}\tabularnewline
\hline
$\phi_{e}/\phi_{g}$ & See Eq.~\eqref{eq:FitPhiEPhiG} \tabularnewline
\hline
$\phi_{s}$ & See Eq.~\eqref{eq:FitPhiS} \tabularnewline
\hline
$\delta_{m0}$ & 310/$K_{x}^{-1}$ \citep{Maloney1973}\tabularnewline
\hline
\end{tabular}
\par\end{centering}
\end{table}

\subsection{The effect of the diffusive influx and carrier efflux}

All the parameters required to study the effect of the diffusive influx
and carrier efflux are now available (Table~\ref{tab:Parameters}).
Most of the simulations shown below were done with the parameter values
for TMG. However, at the end of this section, we shall show the simulation
for IPTG, and compare it with the experimental data.

We shall begin by considering the base case of no diffusive influx
or carrier efflux. To this base case, we shall add the diffusive influx
and the carrier efflux, one at a time, to examine their effects on
the steady states and the thresholds. Finally, we shall show that
that the general model containing both fluxes is essentially a composite
of the two special cases accounting for the diffusive influx and carrier
efflux.

\subsubsection{No diffusive influx or carrier influx}

In the absence of the diffusive influx and the carrier efflux, the
equilibrium condition \eqref{eq:EqCondn} becomes\begin{equation}
\chi=\delta_{m}(G)f(\chi)\frac{\sigma}{\kappa_{1}+\sigma}.\label{eq:Case1EqCondn}\end{equation}
We can determine the variation of the steady states with $\sigma$
at any given $G$ by solving this equation for $\chi$, and substituting
it in \eqref{eq:eSS} and \eqref{eq:gSS}. However, this approach
can only be implemented numerically. It is more convenient to solve
\eqref{eq:Case1EqCondn} for $\sigma$, thus obtaining\begin{equation}
\sigma(\chi,G)=\kappa_{1}\frac{\chi}{\delta_{m}(G)f(\chi)-\chi}.\label{eq:Case1SigmaSS}\end{equation}
The variation of the steady state $\epsilon$, $\gamma$, and $\chi$
with $\sigma$ is then given by the parametric curves, $\left\{ \sigma(\chi,G),\epsilon(\chi,G)\right\} $,
$\left\{ \sigma(\chi,G),\gamma(\chi,G)\right\} $, and $\left\{ \sigma(\chi,G),\chi)\right\} $,
respectively, where $\epsilon(\chi,G)$, $\gamma(\chi,G)$, and $\sigma(\chi,G)$
are given by \eqref{eq:eSS}, \eqref{eq:gSS}, and \eqref{eq:Case1SigmaSS},
respectively.

\begin{figure}[t]
\noindent \begin{centering}
\subfigure[]{\includegraphics[width=2.5in]{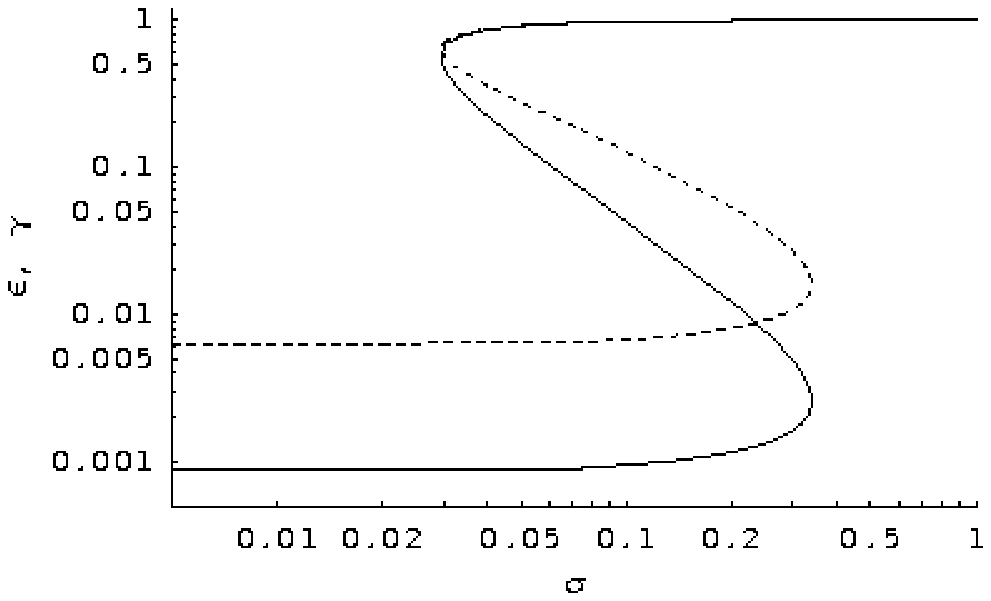}}\hspace*{0.2in}\subfigure[]{\includegraphics[width=2.5in]{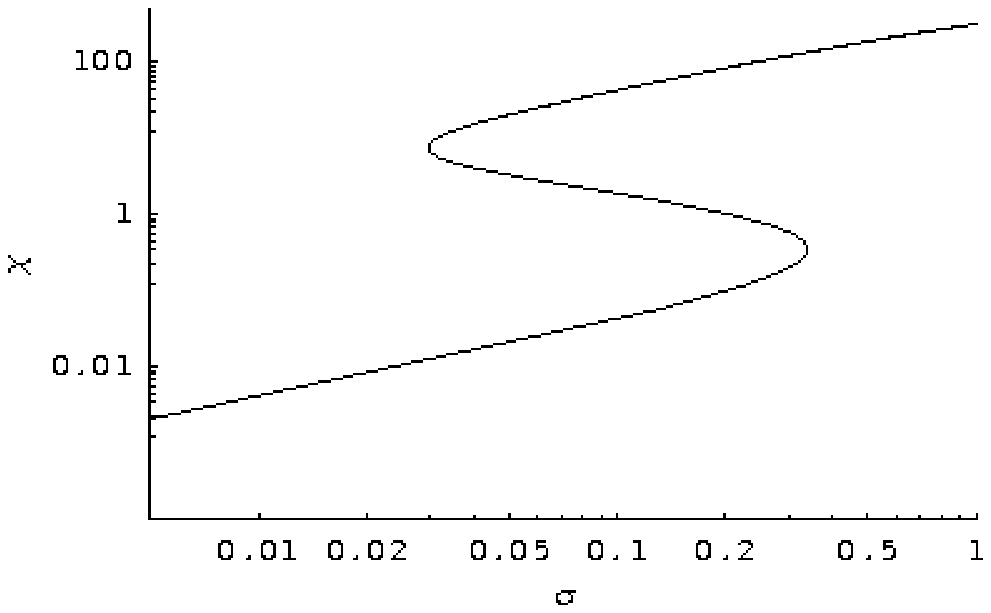}}
\par\end{centering}

\caption{\label{fig:Case1SS}Variation of the steady states with the extracellular
TMG level, $\sigma$, at $G=0$ in the absence of diffusive influx
and carrier efflux. (a)~Enzyme activity (full line) and green fluorescence
intensity (dashed line). (b)~Intracellular TMG level.}

\end{figure}

Figure~\ref{fig:Case1SS}a shows the variation of the steady state
$\epsilon$ and $\gamma$ with $\sigma$ at $G=0$. Both variables
are bistable over the very same $\sim$10-fold range of extracellular
TMG concentrations. Furthermore, the GFP intensity and enzyme activity
of the induced cells are identical ($\epsilon,\gamma\approx1$), but
the GFP intensity of non-induced cells is significantly higher than
their enzyme activity.

The foregoing trends can be understood in terms of the variation of
the intracellular TMG levels with $\sigma$ (Fig.~\ref{fig:Case1SS}b).
In non-induced cells, the GFP intensity is significantly higher than
the enzyme activity because the intracellular TMG levels are so small
compared to $K_{x}^{-1}$ that the reporter and native \emph{lac}
operons are transcribed at their basal rates. Since the reporter \emph{lac}
operon lacks $O_{2}$, its basal transcription rate, $1/(1+\alpha+\hat{\alpha}_{G})=1/170$,
is $\sim$7 times higher than that of the native \emph{lac} operon,
$1/(1+\alpha+\hat{\alpha})=1/1241$. In induced cells, on the other
hand, the inducer levels are so high that the repressor is completely
inactivated, and both operons are transcribed at the very same (maximal)
rates. Finally, the thresholds for $\epsilon$ and $\gamma$ are the
same because \eqref{eq:eSS} and \eqref{eq:gSS} imply that $\epsilon$
and $\gamma$ are increasing functions of $\chi$. Hence\[
\frac{d\sigma}{d\epsilon}=\frac{d\sigma}{d\chi}\frac{d\chi}{d\epsilon},\;\frac{d\sigma}{d\gamma}=\frac{d\sigma}{d\chi}\frac{d\chi}{d\gamma}\]
are zero precisely when $d\sigma/d\chi=0$. Since $d\sigma/d\chi$
is zero at $\sigma\approx0.03$ and $\sigma\approx0.3$, so are $d\sigma/d\epsilon$
and $d\sigma/d\gamma$.

\begin{figure}[t]
\noindent \begin{centering}
\subfigure[]{\includegraphics[width=2.5in]{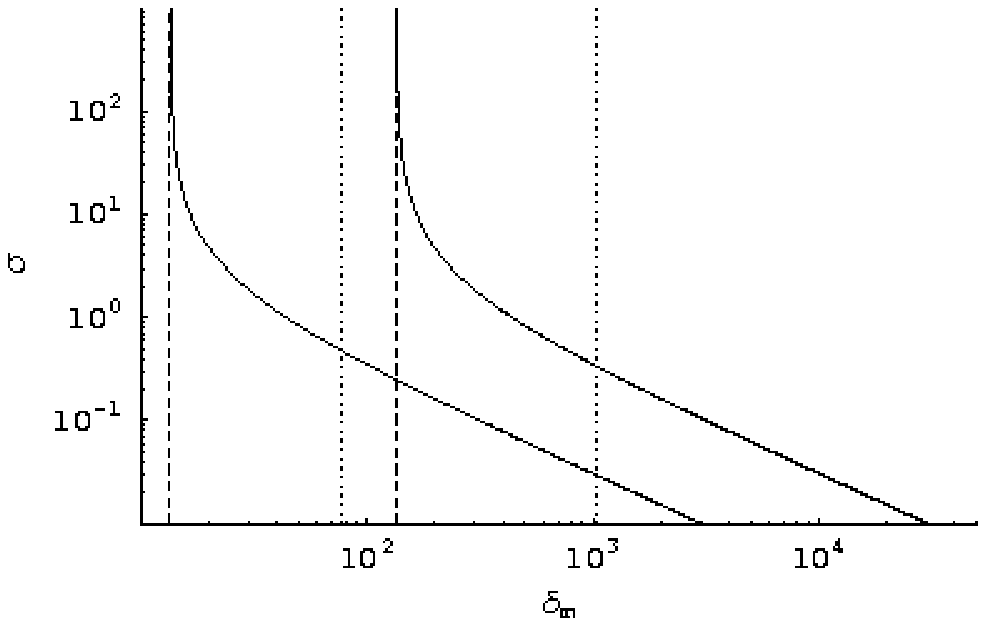}}\hspace*{0.2in}\subfigure[]{\includegraphics[width=2.5in]{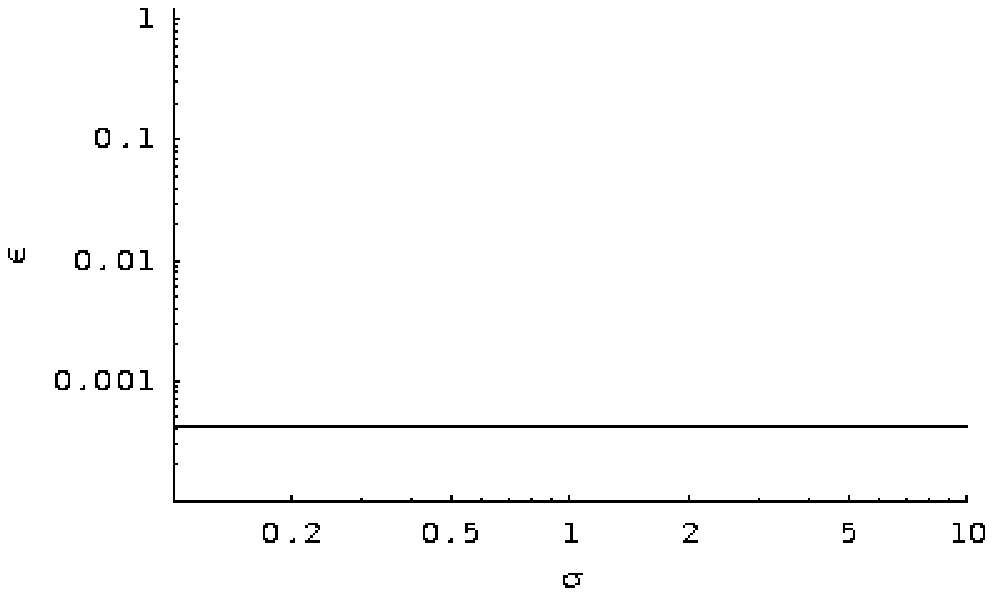}}
\par\end{centering}

\noindent \begin{centering}
\subfigure[]{\includegraphics[width=2.5in]{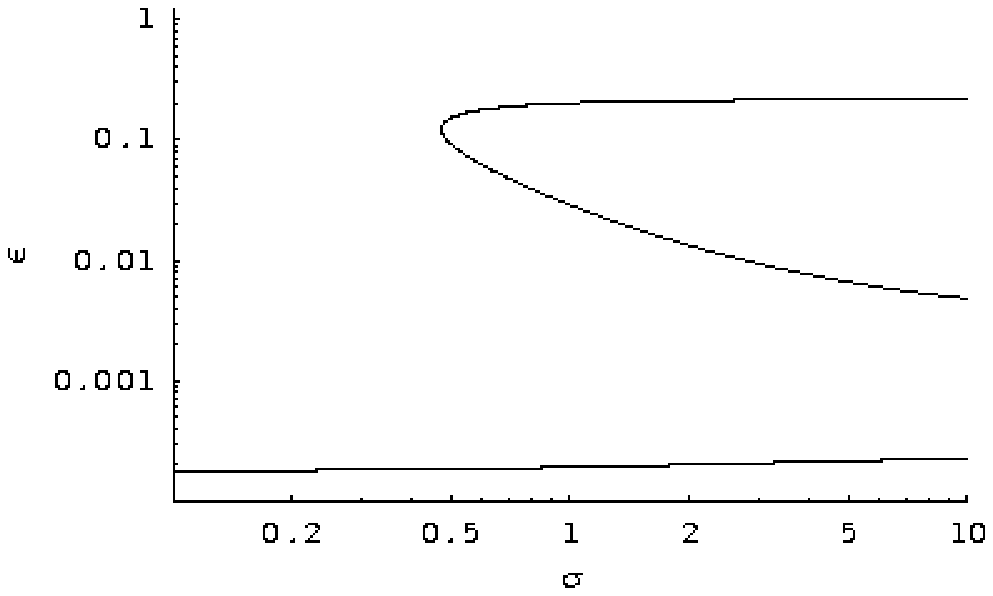}}\hspace*{0.2in}\subfigure[]{\includegraphics[width=2.5in]{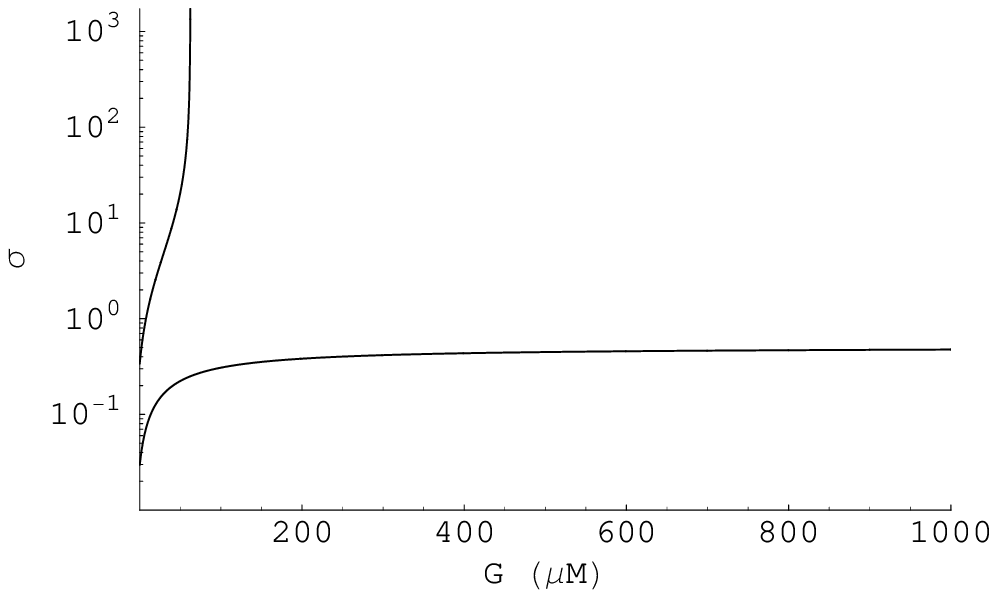}}
\par\end{centering}

\caption{\label{fig:Case1BD}Variation of the on and off thresholds in the
absence of diffusive influx and carrier efflux. (a)~Variation of
the on threshold (upper full curve) and off threshold (lower full
curve) with $\delta_{m}$. The dashed vertical lines represent the
values of $\delta_{m}$ at which the thresholds become infinitely
large. The dotted vertical lines represent the values of $\delta_{m}$
at $G=0$ and $G=1000$~$\mu$M. (b)~If $\delta_{m}=10<1/f_{\chi}(\chi_{2})$,
there is no bistability. (c)~If $\delta_{m}=80$, which lies between
$1/f_{\chi}(\chi_{1})$ and $1/f_{\chi}(\chi_{2})$, there is bistability,
but no on threshold. (d)~Variation of the off threshold (lower curve)
and on threshold (upper curve) with the glucose concentration, $G$.}

\end{figure}

We are particularly interested in the variation of the on and off
thresholds with $G$, the concentration of extracellular glucose.
In this regard, it is worth noting two points. First, since $\epsilon$
and $\gamma$ always have the same thresholds, it suffices to focus
on either one of these variables. In the simulations that follow,
we present only the enzyme activities. Second, although we are interested
in the variation of the thresholds with $G$, we shall begin by determining
their variation with $\delta_{m}$. As we show below, this immediately
yields their variation with $G$ because $\delta_{m}$ is a monotonically
decreasing function of $G$.

In the absence of the carrier efflux, the bifurcation condition \eqref{eq:BifnCondn}
becomes\begin{equation}
1-\delta_{m}f_{\chi}\frac{\sigma}{\kappa_{1}+\sigma}=0.\label{eq:Case1BifnCondn}\end{equation}
The thresholds satisfy \eqref{eq:Case1EqCondn} and \eqref{eq:Case1BifnCondn},
which imply that\begin{equation}
\chi-\frac{f}{f_{\chi}}=0.\label{eq:Case1ChiEqn}\end{equation}
It is shown in Appendix~\ref{app:AnalysisCase1} that \eqref{eq:Case1ChiEqn}
has exactly two positive roots, $\chi_{1}<\chi_{2}$, and they satisfy
the relation, $f_{\chi}(\chi_{1})<f_{\chi}(\chi_{2})$. The loci of
the on and off thresholds are therefore given by the equations\begin{equation}
\sigma=\frac{\kappa_{1}}{\delta_{m}f_{\chi}(\chi_{1})-1},\;\sigma=\frac{\kappa_{1}}{\delta_{m}f_{\chi}(\chi_{2})-1},\label{eq:Case1BifnCurves}\end{equation}
which define two decreasing curves on the $\delta_{m}\sigma$-plane
(Fig.~\ref{fig:Case1BD}a). Furthermore:

\begin{itemize}
\item The on and off thresholds become infinitely large when $\delta_{m}$
approaches $1/f_{\chi}(\chi_{1})$ and $1/f_{\chi}(\chi_{2})$, respectively
(dashed lines in Fig.~\ref{fig:Case1BD}a).
\item When $\delta_{m}\gg1/f_{\chi}(\chi_{1})>1/f_{\chi}(\chi_{2})$, \eqref{eq:Case1BifnCurves}
reduces to\[
\sigma\approx\left(\frac{\kappa_{1}}{f_{\chi}(\chi_{1})}\right)\frac{1}{\delta_{m}},\;\sigma\approx\left(\frac{\kappa_{1}}{f_{\chi}(\chi_{2})}\right)\frac{1}{\delta_{m}}.\]
Thus, the curves are hyperbolic, and the ratio of the on threshold
to the off threshold equals the constant, $f_{\chi}(\chi_{2})/f_{\chi}(\chi_{1})$.
\end{itemize}
Taken together, these results imply that if $\delta_{m}>1/f_{\chi}(\chi_{1})$,
there is bistability with finite on and off thresholds (Fig.~\ref{fig:Case1SS});
if $\delta_{m}<1/f_{\chi}(\chi_{2})$, there is no bistability (Fig.~\ref{fig:Case1BD}b);
and finally, if $1/f_{\chi}(\chi_{2})<\delta_{m}<1/f_{\chi}(\chi_{1})$,
there is bistability with a finite off threshold, but no on threshold
(Fig.~\ref{fig:Case1BD}c). We shall elaborate on the last case shortly.

The variation of the off and on thresholds with the glucose level,
$G$, follows immediately from Fig.~\ref{fig:Case1BD}a. Indeed,
as $G$ increases from 0 to 1000~$\mu$M, $\delta_{m}\equiv\delta_{m0}\phi$
decreases from $\delta_{m0}$ to $\delta_{m\infty}\equiv\left.\delta_{m0}\phi\right|_{G=1000}$,
shown by the dotted vertical lines in Fig.~\ref{fig:Case1BD}a. It
follows that as $G$ increases, the on and off thresholds as well
as the ratio of the on to off threshold increase. Furthermore, the
off threshold exists for all $0\le G\le1000$~$\mu$M, whereas the
on threshold ceases to exist for some $G<1000$~$\mu$M.

We can get explicit expressions for the variation of the thresholds
with $G$ by letting $\delta_{m}=\delta_{m0}\phi(G)$ in \eqref{eq:Case1BifnCurves},
thus obtaining\[
\sigma=\frac{\kappa_{1}}{\delta_{m0}\phi(G)f_{\chi}(\chi_{1})-1},\;\sigma=\frac{\kappa_{1}}{\delta_{m0}\phi(G)f_{\chi}(\chi_{2})-1}.\]
Since $\phi$ decreases with $G$, $\sigma$ increases with $G$.
Thus, on the $G\sigma$-plane, the thresholds are represented by two
increasing curves (Fig.~\ref{fig:Case1BD}d). As expected, these
curves diverge with increasing glucose levels, with the on threshold
becoming unbounded at $G\approx100$~$\mu$M.

The disappearance of the on threshold at intermediate values of $\delta_{m}$
is an example of \emph{irreversible bistability}~\citep[p.~421]{Laurent1999}.
Indeed, it is evident from Fig.~\ref{fig:Case1BD}c that synthesis
of the enzymes in induced cells can be switched off by gradually decreasing
the extracellular inducer concentration. However, once the enzyme
synthesis has been switched off in this manner, it cannot switched
on again by any gradual increase of the extracellular inducer concentration.
Irreversible bistability is thought to play an important role in development
because it furnishes a mechanism for changing the cellular phenotype
permanently. However, it is biologically implausible in the case of
the \emph{lac} operon. Due to the existence of the diffusive influx,
non-induced (and in fact, even cryptic cells) can always be induced
by exposing them to sufficiently high extracellular inducer levels
(Fig.~\ref{fig:Herzenberg}a). We show below that irreversible bistability
disappears in the presence of the diffusive influx.

\subsubsection{Diffusive influx, but no carrier efflux}

In the absence of the carrier efflux, \eqref{eq:EqCondn} becomes\begin{equation}
\chi=\sigma+\delta_{m}(G)f\frac{\sigma}{\kappa_{1}+\sigma},\label{eq:Case2EqCondn}\end{equation}
which can be solved for $\sigma$ to obtain\begin{equation}
\sigma(\chi,G)=\frac{1}{2}\left[-\left(\kappa_{1}+\delta_{m}f-\chi\right)+\sqrt{\left(\kappa_{1}+\delta_{m}f-\chi\right)^{2}+4\kappa_{1}\chi}\right].\label{eq:Case2SigmaSS}\end{equation}
The variation of the steady state $\epsilon$, $\gamma$, and $\chi$
with $\sigma$ is given by the parametric curves, $\left\{ \sigma(\chi,G),\epsilon(\chi,G)\right\} $
and $\left\{ \sigma(\chi,G),\chi)\right\} $, respectively, where
$\epsilon(\chi,G)$ and $\sigma(\chi,G)$ are given by \eqref{eq:eSS}
and \eqref{eq:Case2SigmaSS}, respectively.

\begin{figure}[t]
\noindent \begin{centering}
\subfigure[]{\includegraphics[width=2.5in]{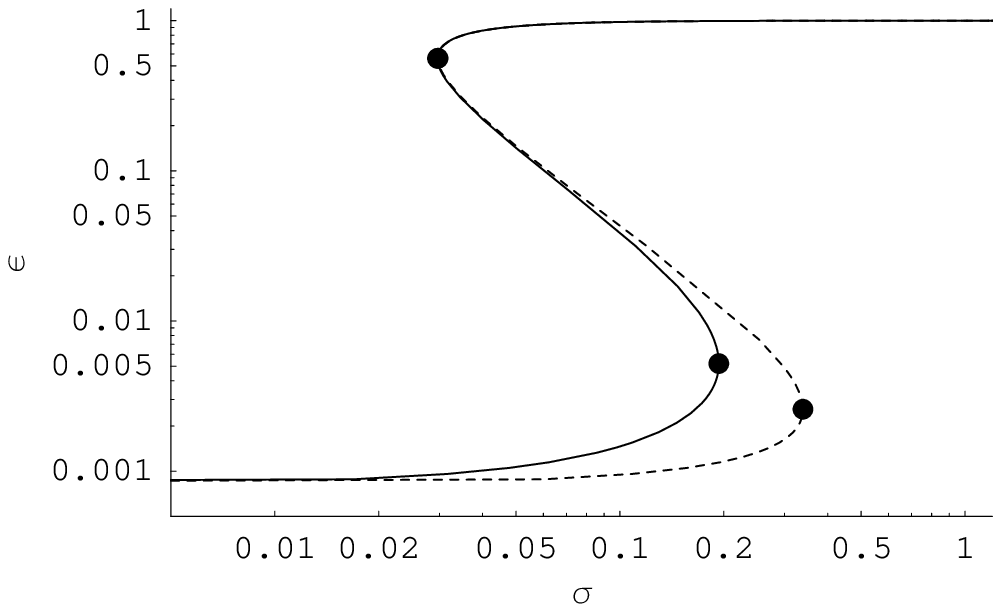}}\hspace*{0.2in}\subfigure[]{\includegraphics[width=2.5in]{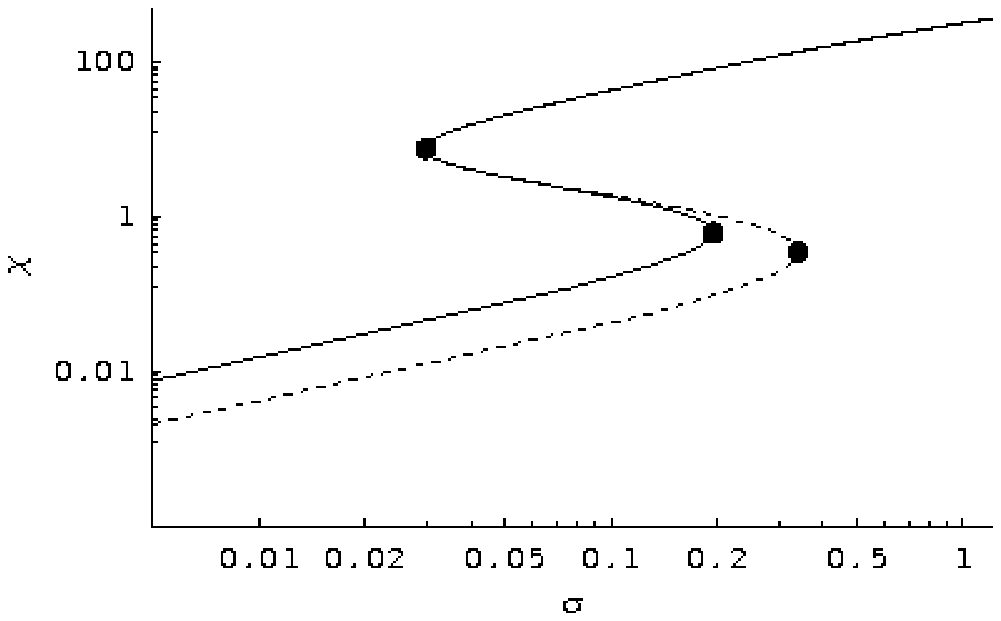}}
\par\end{centering}

\caption{\label{fig:Case2SS}Comparison of the steady states at $G=0$ in the
presence (full lines) and absence (dashed lines) of the diffusive
influx. (a)~Variation of the steady state permease activity, $\epsilon$,
with the extracellular TMG level, $\sigma$. (b)~Variation of the
steady state intracellular TMG concentration, $\chi$, with the extracellular
TMG level, $\sigma$. }

\end{figure}

Figure~\ref{fig:Case2SS}a compares the variation of the steady state
permease activities with $\sigma$ at $G=0$ in the presence (full
curve) and absence (dashed curve) of the diffusive flux. It is evident
that the diffusive flux has no effect on the off threshold, but it
significantly reduces the on threshold. The physical reason for this
is as follows. The off threshold is a property of the induced cells.
These cells contain such high permease levels that the diffusive influx
makes virtually no contribution to the accumulation of TMG (see upper
branches of the curves in Fig.~\ref{fig:Case2SS}b). In contrast,
the on threshold is a property of the non-induced cells, which contain
such low enzyme levels that the diffusive influx significantly improves
the accumulation of TMG (see lower branches of the curves in Fig.~\ref{fig:Case2SS}b).
This enhanced accumulation of TMG decreases the on threshold.

\begin{figure}[t]
\noindent \begin{centering}
\subfigure[]{\includegraphics[width=2.5in]{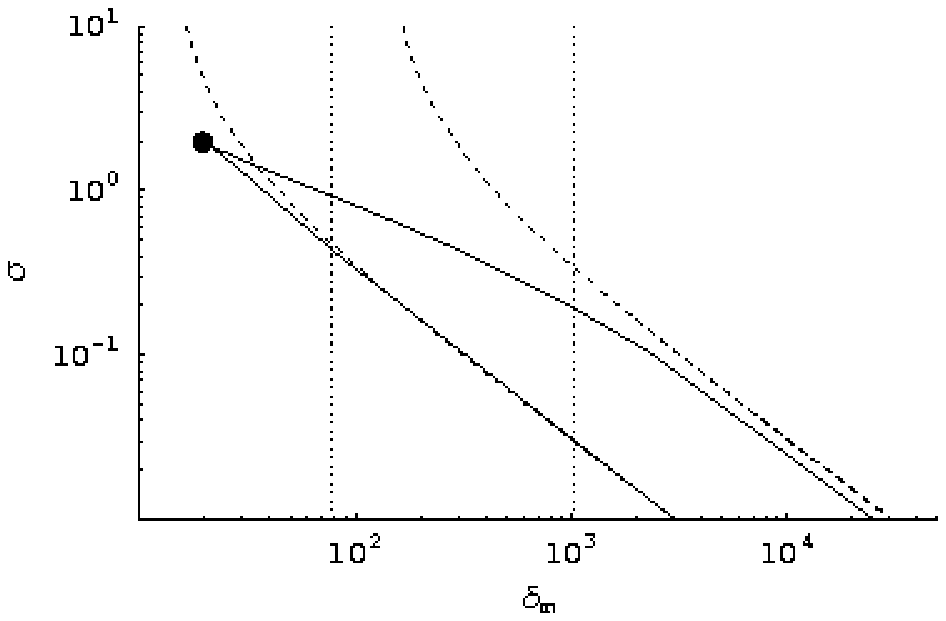}}\hspace*{0.2in}\subfigure[]{\includegraphics[width=2.5in]{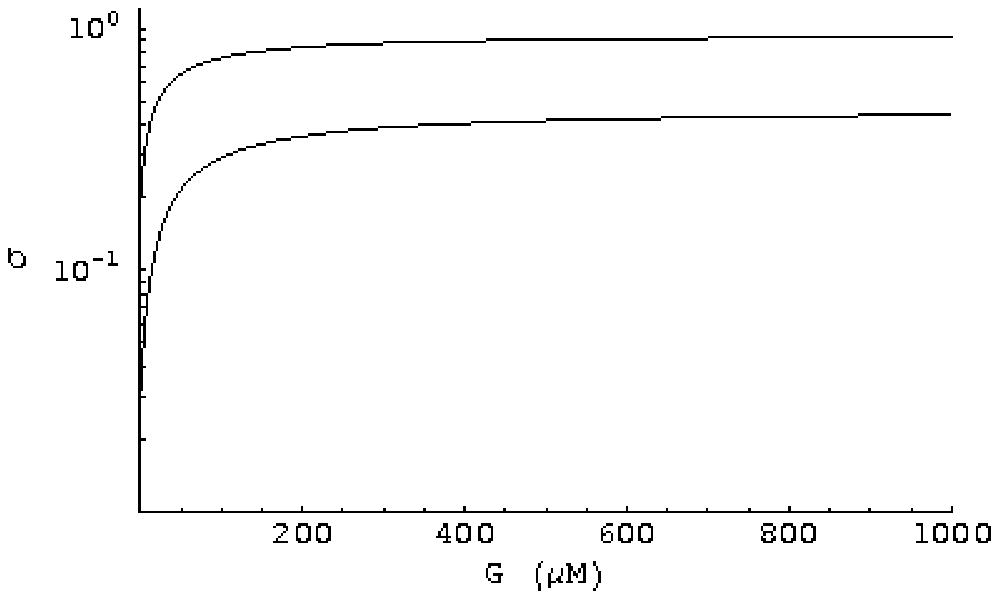}}
\par\end{centering}

\caption{\label{fig:Case2BD}Variation of the on and off thresholds in the
presence of the diffusive influx. (a)~Variation of the on and off
thresholds with $\delta_{m}$. The lower and upper full curves, defined
by \eqref{eq:Case2_Sigma} and \eqref{eq:Case2_DeltaM}, represent
the off and on thresholds, respectively, in the presence of the diffusive
influx. The full circle shows the cusp at which the thresholds merge.
The lower and upper dashed curves, defined by \eqref{eq:Case1BifnCurves},
represent the off and on thresholds, respectively, in the absence
of the diffusive influx. The dotted vertical lines show the values
of $\delta_{m}$ at $G=0$ and $G=1000$~$\mu$M. (b)~Variation
of the off threshold (lower curve) and on threshold (upper curve)
with the glucose concentration, $G$.}

\end{figure}

We have shown above that in the absence of glucose, the diffusive
influx influences only the on threshold. Is this true even in the
presence of glucose? We can address this question by constructing
the loci of the on and off thresholds on the $\delta_{m}\sigma$-plane.
To this end, observe that the thresholds satisfy the equilibrium condition
\eqref{eq:Case2EqCondn} and the bifurcation condition \eqref{eq:Case1BifnCondn}.
These two equations can be solved for $\sigma$ and $\delta_{m}$
to obtain the relations\begin{align}
\sigma(\chi) & =\chi-\frac{f}{f_{\chi}},\label{eq:Case2_Sigma}\\
\delta_{m}(\chi) & =\frac{1}{f_{\chi}}\left[1+\frac{\kappa_{1}}{\sigma(\chi)}\right],\label{eq:Case2_DeltaM}\end{align}
which provide the parametric representation, $\left\{ \delta_{m}(\chi),\sigma(\chi)\right\} $,
of the thresholds on the $\delta_{m},\sigma$-plane. The full curve
in Fig.~\ref{fig:Case2BD}a shows a plot of this representation.
It consists of an upper branch and a lower branch, which represent
the on and off thresholds, respectively. Since these two branches
meet in a cusp, bistability is reversible whenever it exists: Either
both or none of the thresholds occur at any given $\delta_{m}$. As
expected, irreversible bistability, which corresponds to the existence
of only one of the thresholds, disappears in the presence of the diffusive
flux. Importantly, this conclusion does not depend on the precise
values of the parameter values chosen for the simulations. It is shown
in Appendix~\ref{app:AnalysisCase2} that the geometry of the curves
representing the thresholds is preserved, regardless of the parameter
values.

To address the question posed above, it is useful to compare the thresholds
in the presence of the diffusive influx with those obtained in its
absence (dashed curves in Fig.~\ref{fig:Case2BD}a). This comparison
shows that there are 3 distinct regimes. When $\delta_{m}\gtrsim2000$,
the full and dashed curves coincide, i.e., the diffusive influx has
no effect on both thresholds. At large $\delta_{m}$, the permease
is so active that even in the non-induced cells, the diffusive influx
makes no contribution to inducer accumulation. When $100\lesssim\delta_{m}$$\lesssim$2000,
the lower full and dashed curves coincide, but the upper full curve
is significantly lower than the upper dashed curve. The influence
of the diffusive influx is therefore identical to that observed at
$G=0$: It has no effect on the off threshold, and significantly decreases
the on threshold. At these intermediate $\delta_{m}$, the enzyme
activity of the induced cells remains so large that the diffusive
influx makes no contribution to inducer accumulation, but the enzyme
activity of the non-induced cells becomes so small that the diffusive
influx does enhance inducer accumulation. Finally, when $\delta_{m}\lesssim100$,
the enzyme activity becomes so small that the diffusive flux contributes
significantly to inducer accumulation in both non-induced and induced
cells. In fact, at sufficiently small $\delta_{m}$, the inducer accumulates
almost entirely by diffusion, i.e., $\chi\approx\sigma$. Since the
intracellular TMG level is independent of the enzyme level, there
is no positive feedback, and hence, no bistability. This is manifested
in Fig.~\ref{fig:Case2BD}a by the formation of a cusp at which the
two thresholds merge.

The qualitative variation of the thresholds with the glucose concentration
can be inferred from Fig.~\ref{fig:Case2BD}a. Indeed, as the glucose
concentration is increased from 0 to 1000~$\mu$M, $\delta_{m}$
decreases from $\delta_{m0}$ to $\delta_{m\infty}$, shown by the
dotted lines in the figure. Evidently, as $G$ increases, both thresholds
increase, but the ratio of the on to off threshold decreases (in contrast
to the increase observed in the absence of diffusive flux).

The quantitative variation of the thresholds with $G$ is obtained
by observing that \eqref{eq:Case2_DeltaM} yields\begin{equation}
\phi(G)=\frac{1}{\delta_{m0}f_{\chi}}\left[1+\frac{\kappa_{1}}{\sigma(\chi)}\right]\Rightarrow G(\chi)=\phi^{-1}\left[\frac{1}{\delta_{m0}f_{\chi}}\left\{ 1+\frac{\kappa_{1}}{\sigma(\chi)}\right\} \right].\label{eq:Case2G}\end{equation}
The locus of the thresholds on the $G\sigma$-plane is therefore given
by the parametric representation, $\left\{ G(\chi),\sigma(\chi)\right\} $,
where $G(\chi)$ and $\sigma(\chi)$ are given by \eqref{eq:Case2G}
and \eqref{eq:Case2_Sigma}, respectively. Figure~\ref{fig:Case2BD}b
shows that this representation yields two increasing curves that approach
each other as $G$ increases.

\subsubsection{Carrier efflux, but no diffusive influx}

In this case, \eqref{eq:EqCondn} becomes\begin{equation}
\chi=\frac{\delta_{m}(G)f(\chi)\sigma/(\kappa_{1}+\sigma)}{1+\delta_{m}(G)f(\chi)/\kappa_{2}},\label{eq:Case3EqCondn}\end{equation}
which can be solved for $\sigma$ to obtain\begin{equation}
\sigma(\chi,G)=\kappa_{1}\frac{\chi\left[1/(\delta_{m}f)+1/\kappa_{2}\right]}{1-\chi\left[1/(\delta_{m}f)+1/\kappa_{2}\right]}.\label{eq:Case3SigmaSS}\end{equation}
The variation of the steady state $\epsilon$ and $\chi$ with $\sigma$
is given by the parametric curves, $\left\{ \sigma(\chi,G),\epsilon(\chi,G)\right\} $
and $\left\{ \sigma(\chi,G),\chi)\right\} $, respectively, where
$\epsilon(\chi,G)$ and $\sigma(\chi,G)$ are given by \eqref{eq:eSS}
and \eqref{eq:Case3SigmaSS}, respectively.

\begin{figure}[t]
\noindent \begin{centering}
\subfigure[]{\includegraphics[width=2.5in]{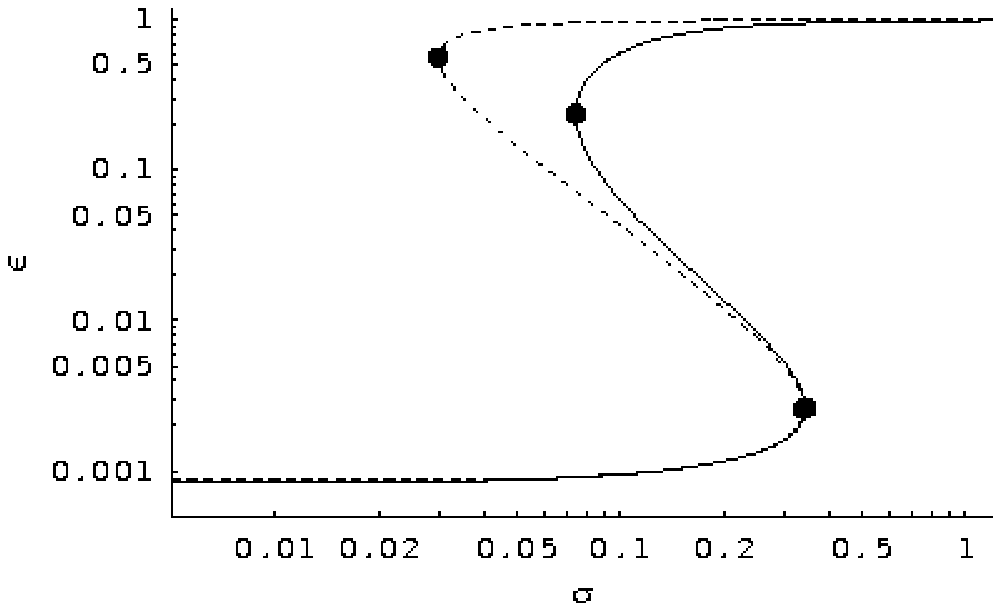}}\hspace*{0.2in}\subfigure[]{\includegraphics[width=2.5in]{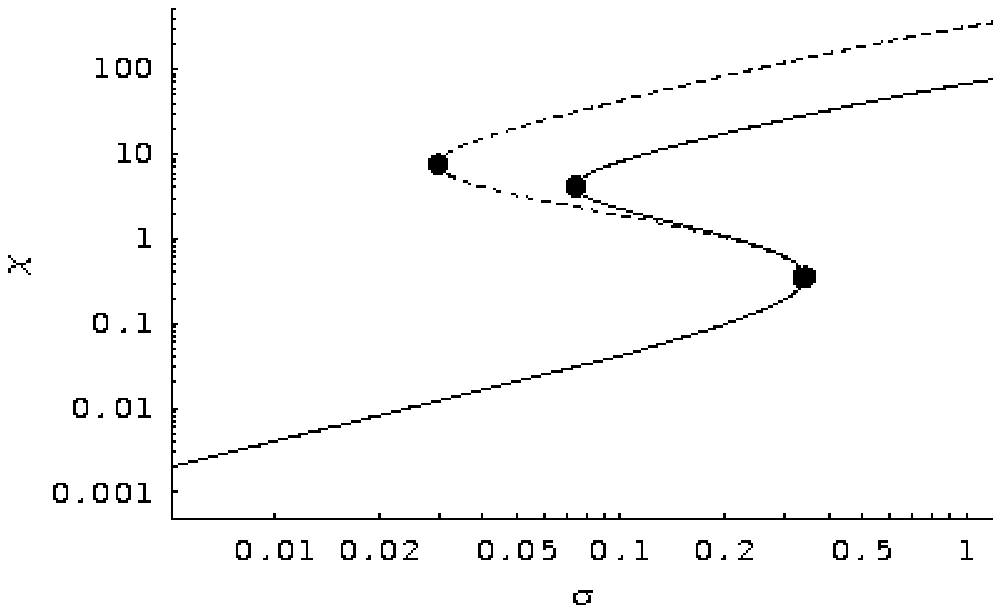}}
\par\end{centering}

\caption{\label{fig:Case3SS}Comparison of the steady states at $G=0$ in the
presence (full lines) and absence (dashed lines) of the carrier efflux.
(a)~Variation of the steady state enzyme activity with the extracellular
TMG level. (b)~Variation of the intracellular TMG concentration with
the extracellular TMG level. }

\end{figure}

Figure~\ref{fig:Case3SS}a shows the variation of the enzyme activity
with $\sigma$ at $G=0$ in the presence (full curve) and absence
(dashed curve) of the carrier influx. Evidently, the carrier efflux
has no effect on the on threshold, but it significantly increases
the off threshold. This is because $\kappa_{2}$ is so large that
the carrier efflux is negligible in non-induced cells, which contain
relatively small enzyme and inducer levels (Fig.~\ref{fig:Case3SS}b).
However, the carrier efflux significantly reduces the the intracellular
TMG level of the induced cells. The net effect of this reduction is
to increase the off threshold of the induced cells.

It is instructive once again to study the variation of the thresholds
with $\delta_{m}$. The thresholds satisfy the equilibrium condition
\eqref{eq:Case3EqCondn} and the bifurcation condition \eqref{eq:BifnCondn},
which can be solved for $\delta_{m}$ and $\sigma$ to obtain\begin{align}
\delta_{m}(\chi) & =\frac{\kappa_{2}}{f}\left(\frac{\chi f_{\chi}}{f}-1\right),\label{eq:Case3_DeltaM}\\
\sigma(\chi) & =\kappa_{1}\frac{\chi^{2}f_{\chi}}{\delta_{m}(\chi)f^{2}-\chi^{2}f_{\chi}}.\label{eq:Case3_Sigma}\end{align}
These relations define the parametric representation, $\left\{ \delta_{m}(\chi),\sigma(\chi)\right\} $,
of the bifurcation curve on the $\delta_{m}\sigma$-plane.

\begin{figure}[t]
\noindent \begin{centering}
\subfigure[]{\includegraphics[width=2.5in]{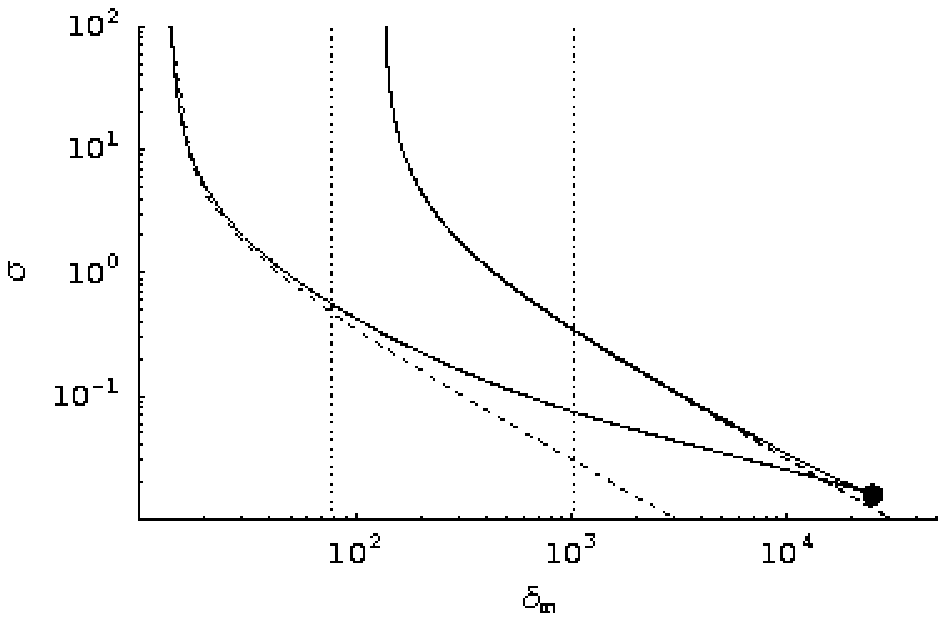}}\hspace*{0.2in}\subfigure[]{\includegraphics[width=2.5in]{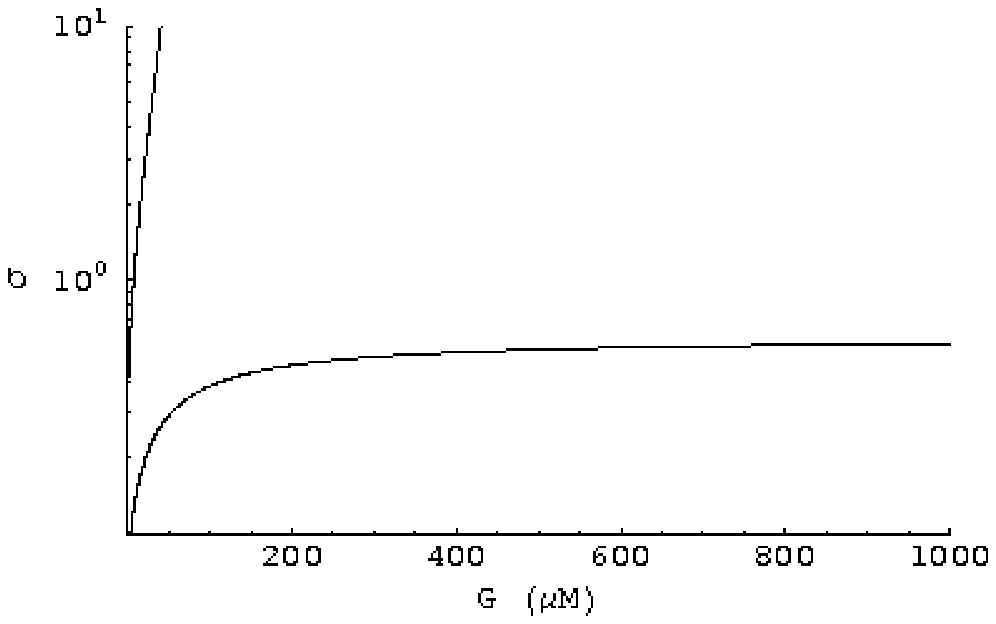}}
\par\end{centering}

\caption{\label{fig:Case3BD}Variation of the on and off thresholds in the
presence of the carrier efflux. (a)~Variation of the on and off thresholds
with $\delta_{m}$. The lower and upper full curves, defined by \eqref{eq:Case3_DeltaM}
and \eqref{eq:Case3_Sigma}, represent the off and on thresholds,
respectively, in the presence of the carrier efflux. The full circle
shows the cusp at which the thresholds merge. The lower and upper
dashed curves, defined by \eqref{eq:Case1BifnCurves}, represent the
off and on thresholds, respectively, in the absence of the carrier
efflux. The dotted vertical lines show the values of $\delta_{m}$
at $G=0$ and $G=1000$~$\mu$M. (b)~Variation of the off threshold
(lower curve) and on threshold (upper curve) with the glucose concentration,
$G$.}

\end{figure}

Figure~\ref{fig:Case3BD}a compares the thresholds in the presence
of carrier efflux (full curves) with those obtained in its absence
(dashed curves). Evidently, when $\delta_{m}\lesssim100$, the carrier
efflux has no effect on both thresholds. Under this condition, the
enzyme activity and the intracellular TMG level are so small that
carrier efflux is negligible even in the induced cells. As $\delta_{m}$
increases, carrier efflux from the induced cells becomes more significant,
and the off threshold becomes progressively larger than that predicted
in the absence of the carrier flux. When $\delta_{m}\gtrsim10^{4}$,
the enzyme activity is so large that even the non-induced cells are
subject to carrier efflux. At sufficiently large $\delta_{m}$, the
off threshold merges with the on threshold at a cusp, beyond which
there is no bistability. The bistability disappears because under
this condition, the intracellular TMG level is\[
\chi\approx\kappa_{2}\frac{\sigma}{\kappa_{1}+\sigma},\]
which is independent of the enzyme level. The destabilizing effect
of positive feedback therefore vanishes, and the prospect of bistability
is eliminated.

Once again, the qualitative variation of the thresholds with $G$
can be inferred from Fig.~\ref{fig:Case3BD}a. As $G$ increases
from 0 to 1000~$\mu$M, $\delta_{m}$ decreases from $\delta_{m0}$
to $\delta_{m\infty}$, shown by the dotted lines in the figure. Evidently,
as $G$ increases, so do the on and off thresholds, as well as the
ratio of the on to off threshold. Furthermore, since diffusive influx
is absent, there is irreversible bistability: The off threshold exists
for all $G$, whereas the on threshold disappears at some $G<1000$~$\mu$M.

The quantitative variation of the thresholds with $G$ follows from
\eqref{eq:Case3_DeltaM}, which yields\begin{equation}
\phi(G)=\frac{\kappa_{2}}{\delta_{m0}f}\left(\frac{\chi f_{\chi}}{f}-1\right)\Rightarrow G(\chi)=\phi^{-1}\left[\frac{\kappa_{2}}{\delta_{m0}f}\left(\frac{\chi f_{\chi}}{f}-1\right)\right].\label{eq:Case3G}\end{equation}
The variation of the thresholds with $G$ is given by the parametric
representation, $\left\{ G(\chi),\sigma(\chi)\right\} $, where $G(\chi)$
and $\sigma(\chi)$ are given by \eqref{eq:Case3G} and \eqref{eq:Case3_Sigma},
respectively. This representation defines two increasing curves departing
from each other (Fig.~\ref{fig:Case3BD}b).

We conclude by observing that the geometry of the thresholds (full
curves in Figure~\ref{fig:Case3BD}a) does not depend on the precise
values of the parameters used in the simulations (see Appendix~\ref{app:AnalysisCase3}
for details).

\subsubsection{Diffusive influx and carrier efflux}

In this case, the steady state intracellular TMG level is given by
\eqref{eq:EqCondn}, which can be solved for $\sigma$ to obtain\begin{equation}
\sigma(\chi,G)=\frac{1}{2}\left[-\left(\kappa_{1}+\delta_{m}f-p\right)+\sqrt{\left(\kappa_{1}+\delta_{m}f-p\right)^{2}+4\kappa_{1}p}\right],\label{eq:Case4SigmaSS}\end{equation}
where $p\equiv\chi\left(1+\delta_{m}f/\kappa_{2}\right)$. The variation
of the steady state $\epsilon$ and $\chi$ with $\sigma$ is given
by the parametric curves, $\left\{ \sigma(\chi,G),\epsilon(\chi,G)\right\} $
and $\left\{ \sigma(\chi,G),\chi)\right\} $, respectively, where
$\epsilon(\chi,G)$ and $\sigma(\chi,G)$ are given by \eqref{eq:eSS}
and \eqref{eq:Case4SigmaSS}, respectively.

\begin{figure}[t]
\noindent \begin{centering}
\subfigure[]{\includegraphics[width=2.5in]{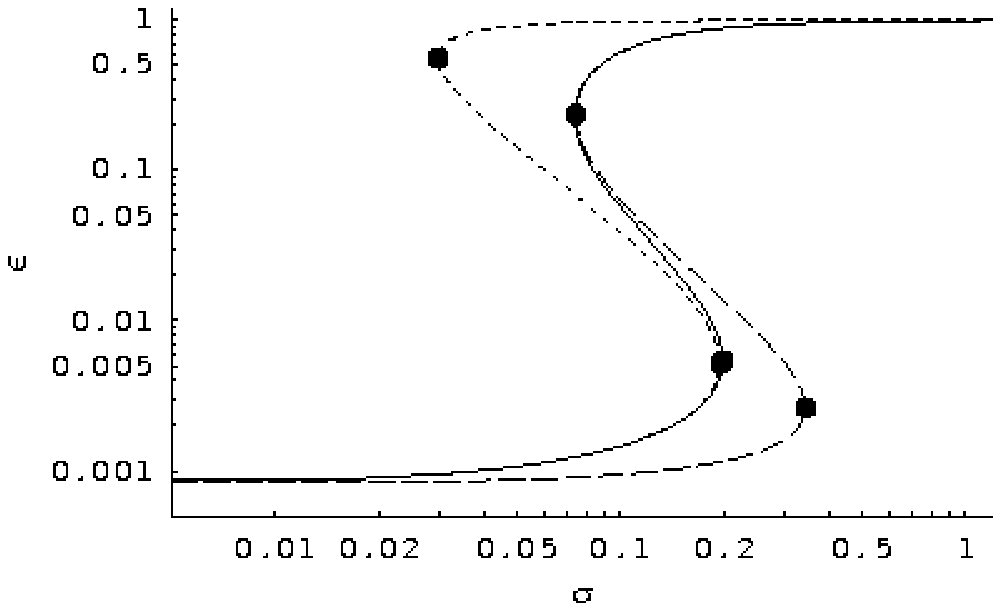}}\hspace*{0.2in}\subfigure[]{\includegraphics[width=2.5in]{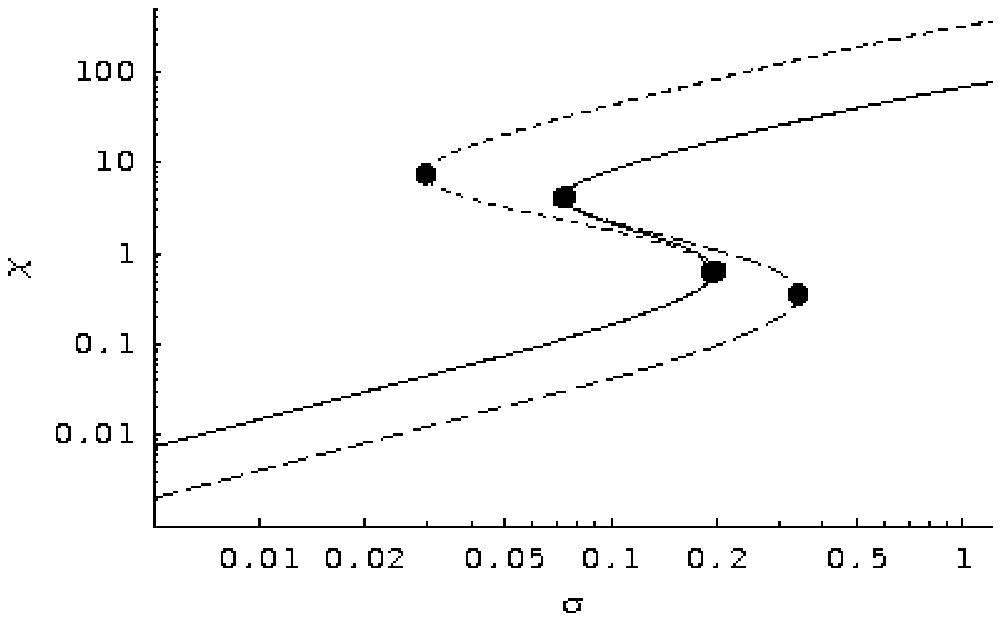}}
\par\end{centering}

\caption{\label{fig:Case4SS}Comparison of the steady states of the general
model (full curves) with the steady states in the presence of diffusive
flux (short-dashed curves) and carrier efflux (long-dashed curves).
(a)~Variation of the enzyme activity with the extracellular TMG level.
(b)~Variation of the intracellular TMG concentration with the extracellular
TMG level. }

\end{figure}

Fig.~\ref{fig:Case4SS} compares the steady state profiles obtained
at $G=0$ in the presence of the diffusive influx and carrier efflux
(full curves) with those obtained in the presence of the diffusive
influx only (short-dashed curves) and the carrier efflux only (long-dashed
curves). It is evident that the upper and lower branches of the steady
state curve, representing the induced and non-induced cells, respectively,
are a composite of the steady states obtained in the presence of the
diffusive influx and carrier efflux. Indeed, the lower branch of the
full curve coincides with the lower branch of short-dashed curve,
and the upper branch of the full curve coincides with the upper branch
of the long-dashed curve. It follows that the steady states and threshold
of the non-induced (resp., induced) cells is well approximated by
the model accounting for only the diffusive flux (resp., carrier efflux).
The reason for this is clear from the above discussion of the special
cases. In the absence of glucose, the diffusive influx influences
the TMG accumulation of non-induced cells only --- it is vanishingly
small in induced cells. Likewise, carrier efflux influences the TMG
accumulation of induced cells only --- it plays no role in non-induced
cells.

The foregoing arguments were based on the steady state profiles obtained
at $G=0$, i.e., $\delta_{m}=\delta_{m0}$. It turns out that they
are true at almost all $\delta_{m}$. This becomes clear if we compare
the thresholds for the general case with the thresholds obtained in
the presence of diffusive influx or carrier efflux. The thresholds
for the general case satisfy \eqref{eq:EqCondn} and \eqref{eq:BifnCondn},
which can be rewritten as\begin{align}
0 & =\sigma^{2}\left[\frac{\chi f_{\chi}+f}{\kappa_{2}}-f_{\chi}\right]+\sigma\left[\left(\chi f_{\chi}-f\right)-\frac{\kappa_{1}\left(f+\chi f_{\chi}\right)-\chi^{2}f_{\chi}}{\kappa_{2}}\right]-\frac{\kappa_{1}\chi^{2}f_{\chi}}{\kappa_{2}},\label{eq:Case4_Sigma}\\
\delta_{m} & =\frac{1}{\frac{\sigma}{\kappa_{1}+\sigma}f_{\chi}-\frac{1}{\kappa_{2}}\left(\chi f_{\chi}+f\right)}.\label{eq:Case4_DeltaM}\end{align}
The first equation can be solved to obtain $\sigma(\chi)$, which
can then be substituted in \eqref{eq:Case4_DeltaM} to obtain $\delta_{m}(\chi)$.
These functions determine the thresholds via the parametric representation,
$\left\{ \delta_{m}(\chi),\sigma(\chi)\right\} $.

\begin{figure}[t]
\noindent \begin{centering}
\subfigure[]{\includegraphics[width=2.5in]{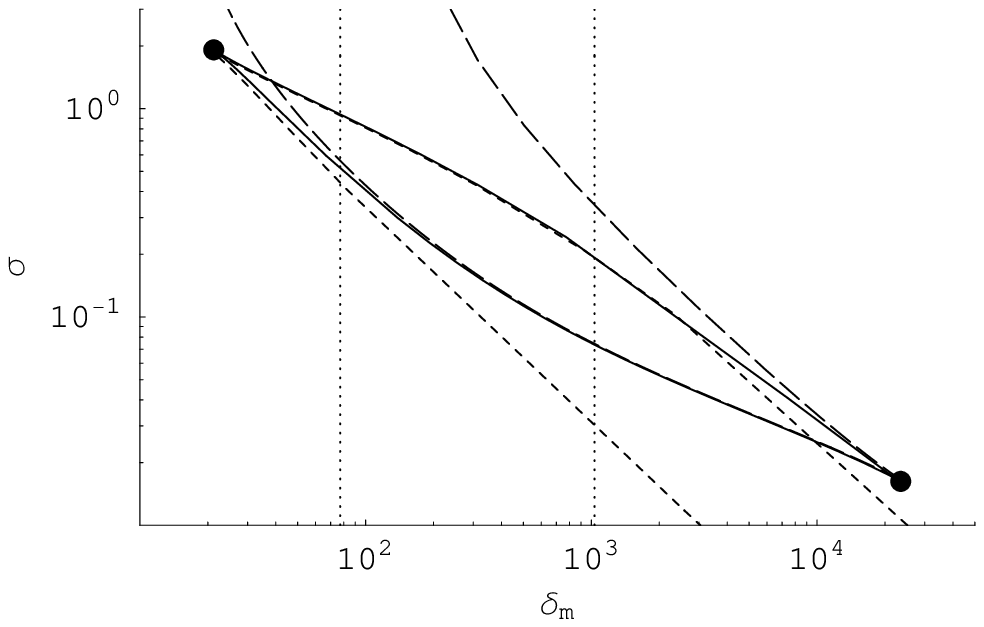}}\hspace*{0.2in}\subfigure[]{\includegraphics[width=2.5in]{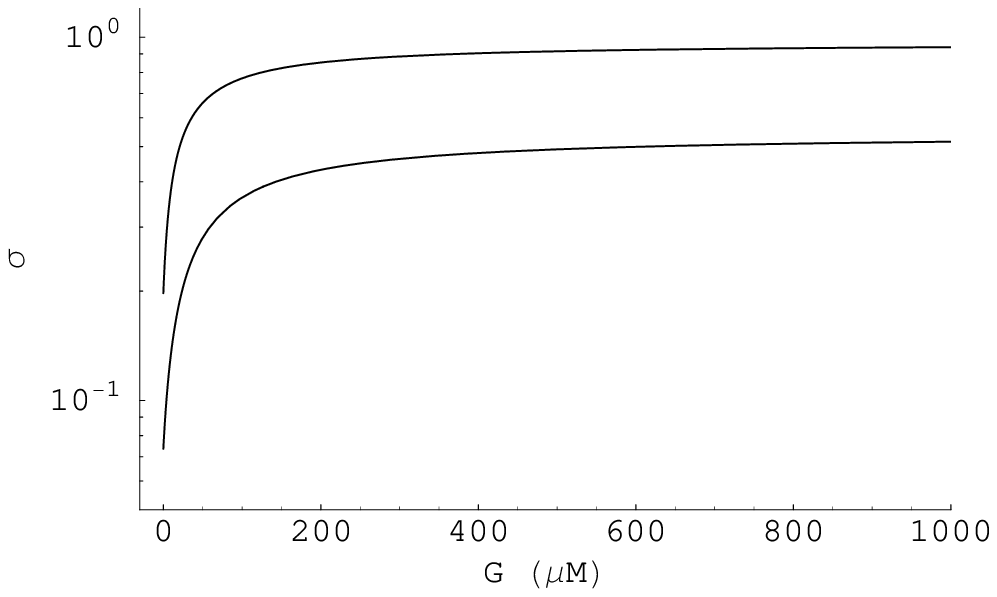}}
\par\end{centering}

\caption{\label{fig:Case4BD}Variation of the on and off thresholds in the
presence of the diffusive influx and the carrier efflux. (a)~Variation
of the on and off thresholds with $\delta_{m}$. The lower and upper
full curves, defined by \eqref{eq:Case4_Sigma} and \eqref{eq:Case4_DeltaM},
represent the off and on thresholds, respectively, in the presence
of diffusive influx and carrier efflux. The full circles show the
cusps at which the thresholds merge. The short-dashed curves, defined
by \eqref{eq:Case2_Sigma} and \eqref{eq:Case2_DeltaM}, represent
the thresholds in the presence of the diffusive influx only. The long-dashed
curves, defined by \eqref{eq:Case3_DeltaM} and \eqref{eq:Case3_Sigma},
represent the thresholds in the presence of the carrier efflux only.
The dotted vertical lines show the values of $\delta_{m}$ at $G=0$
and $G=1000$~$\mu$M. (b)~Variation of the on and off thresholds
with the glucose concentration, $G$.}

\end{figure}

Figure~\ref{fig:Case4BD}a compares the thresholds thus obtained
(full curves) with the thresholds obtained in the presence of diffusive
influx (short-dashed curves) and carrier efflux (long-dashed curves).
The thresholds exist only on a finite interval --- they terminate
in a cusp at both small and large $\delta_{m}$. Over almost its entire
range of existence, the upper branch of the full curve overlaps with
the upper branch of the short-dashed curve, which represents the on
threshold in the presence of the diffusive flux. Likewise, the lower
branch of the full curve coincides with the lower branch of the long-dashed
curve, which represents the off threshold in the presence of the carrier
flux. Thus, the thresholds of the general case are a composite of
the two special cases. The on (resp., off) threshold is well approximated
by \eqref{eq:Case2_Sigma} and \eqref{eq:Case2_DeltaM} {[}resp.,
\eqref{eq:Case3_DeltaM} and \eqref{eq:Case3_Sigma}], which represent
the thresholds in the presence of only the diffusive flux (resp.,
carrier efflux).

It is evident from Fig.~\ref{fig:Case4BD}a that as $G$ increases,
both thresholds increase, but the ratio of on to off thresholds can
increase, decrease, or pass through a maximum. For the parameter values
shown in Table~\ref{tab:Parameters}, $\delta_{m0}$ and $\delta_{m\infty}$
are such that the ratio passes through a maximum. The precise variation
of the thresholds with $G$ is given by the parametric curve, $\left\{ G(\chi),\sigma(\chi)\right\} $,
where \[
G(\chi)=\phi^{-1}\left[\frac{1}{\delta_{m0}\left\{ \frac{\sigma}{\kappa_{1}+\sigma}f_{\chi}-\frac{1}{\kappa_{2}}\left(\chi f_{\chi}+f\right)\right\} }\right],\]
a relation that follows immediately from \eqref{eq:Case4_DeltaM}.
This parametric curve consists of two increasing curves that diverge
at low glucose concentrations, and converge at high glucose concentrations
(Fig.~\ref{fig:Case4BD}b).

\subsubsection{Comparison of the model simulations with the data for TMG}

Comparison of the model simulations (dashed lines in Fig.~\ref{f:Oudenaarden1}b)
with the data shows that the simulated off thresholds are higher,
and the simulated on thresholds are lower, than the corresponding
thresholds observed experimentally. Part of this discrepancy probably
stems from the fact that the parameter values, which correspond to
various strains of \emph{E. coli}, are somewhat different from those
shown in Table~\ref{tab:Parameters}. However, the analysis of the
data suggests that there are two additional sources of the discrepancy.

\begin{figure}[t]
\noindent \begin{centering}
\includegraphics[width=2.5in]{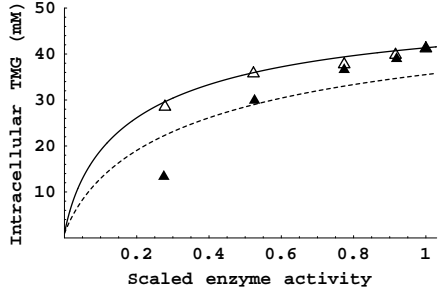}
\par\end{centering}

\caption{\label{fig:InducerExclusion}Comparison of the observed and predicted
effects of inducer exclusion in \emph{S. typhimurium} SB3939 transfected
with a plasmid encoding the \emph{lac} operon~\citep[Fig.~5A]{Mitchell1982}.
The open triangles show the steady state intracellular TMG level in
cells induced to various levels, and exposed to a carbon-free medium
containing 1~mM TMG. The full triangles show the intracellular TMG
level attained if the medium contains 1~mM TMG and 10~mM $\alpha$-methylglucoside.
The full curve shows the fit to the data obtained with Eq.~\eqref{eq:MaloneyXIndEx}
and the parameter values, $\left.\phi_{s}\right|_{G=0}=1$, $k_{x}=0.14$
min$^{-1}$, $s=1$~mM, $K_{1}=0.8$~mM, and $K_{2}=41$~mM. The
dashed curve shows the fit to the data with the same equation and
parameter values, the only difference being that $\left.\phi_{s}\right|_{G=10\textnormal{mM}}=0.5$.}

\end{figure}

\paragraph*{Dependence of $\phi_{s}$ on the the enzyme activity }

The model assumes that the effect of inducer exclusion, characterized
by the function $\phi_{s}$, is completely determined by the extracellular
glucose level. However, the data in Fig.~\ref{fig:InducerExclusion}
implies that $\phi_{s}$ also depends on the activity of the permease.
The open triangles in the figure show the steady state intracellular
TMG levels in cells induced to various levels, and then exposed to
a carbon-free medium containing 1~mM TMG. The full triangles show
the steady state intracellular levels attained when the carbon-free
medium contains 1~mM TMG and 10~mM $\alpha$-methylglucoside ($\alpha$MG),
a non-metabolizable analog of glucose that mimics its inducer exclusion
effect. It is clear from the data that inducer exclusion is significant
in cells containing low enzyme levels, but it disappears in fully
induced cells.

If $\phi_{s}$ is completely determined by the extracellular glucose
concentration, there is no combination of model parameters that can
capture the data in Fig.~\ref{fig:InducerExclusion}. To see this,
observe that the data was obtained in a carbon-free medium, which
prevents efficient enzyme synthesis. The appropriate equation for
modeling the data is therefore given by the following variant of Eq.~\eqref{eq:MaloneyX}
\begin{equation}
0=V\phi_{s}\left(\frac{s}{K_{1}+s}-\frac{x}{K_{2}+x}\right)-k_{x}\left(x-s\right)\Leftrightarrow V=\frac{1}{\phi_{s}(G)}\frac{k_{x}(x-s)}{\frac{s}{K_{1}+s}-\frac{x}{K_{2}+x}}.\label{eq:MaloneyXIndEx}\end{equation}
Since the extracellular TMG level, $s$, is constant (1~mM), the
above equation implies that at any given $x$, the enzyme activity
in the absence of $\alpha$MG (or glucose) is $\phi_{s}(G)$ times
the activity in the presence of $\alpha$MG. This contradicts the
data. At $x\approx30$~mM, for instance, the enzyme activity in the
absence of $\alpha$MG is roughly half the activity in the presence
of 10~mM $\alpha$MG. However, at $x\approx40$~mM, the activities
are the same in the absence and presence of $\alpha$MG.

The molecular explanation for the data in Fig.~\ref{fig:InducerExclusion}
suggests that $\phi_{s}$ depends not only on $G$, but also on the
prevailing enzyme level~\citep{Mitchell1982}. Indeed, inducer exclusion
occurs because the uptake of $\alpha$MG or glucose results in the
formation of dephosphorylated enzyme IIA$^{\textnormal{glc}}$ (a
transport enzyme for glucose), which inhibits \emph{lac} permease
by binding to it. In fully induced cells, the permease level is so
high that most of it is essentially free of IIA$^{\textnormal{glc}}$.
As the permease level decreases, so does the inhibition due to LacY-IIA$^{\textnormal{glc}}$
binding.%
\footnote{No data was obtained at scaled enzyme activities below $\sim0.25$
because the recombinant strain used in these experiments was quasi-constitutive.%
} Thus, the magnitude of the inducer exclusion effect depends on the
prevailing permease level.

The model, which ignores the dependence of $\phi_{s}$ on the permease
level, provides a good fit to the data obtained in the absence of
$\alpha$-MG (full line in Fig.~\ref{fig:InducerExclusion}). However,
it fails to capture the data obtained in the presence of $\alpha$-MG
(dashed line in Fig.~\ref{fig:InducerExclusion}). It overestimates
the inducer exclusion effect in induced cells, and underestimates
this effect in cells containing lower enzyme levels. Consequently,
the observed off threshold is expected to be lower, and the observed
on threshold is expected to be higher, than the corresponding thresholds
predicted by the model.

\begin{figure}[t]
\noindent \begin{centering}
\subfigure[]{\includegraphics[width=2.5in]{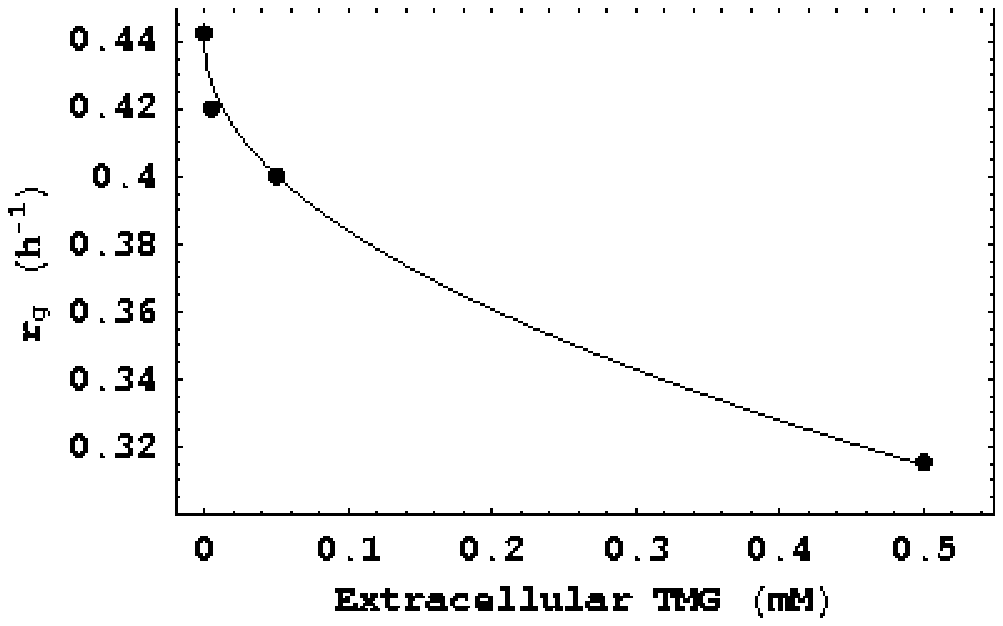}}\hspace*{0.2in}\subfigure[]{\includegraphics[width=2.5in]{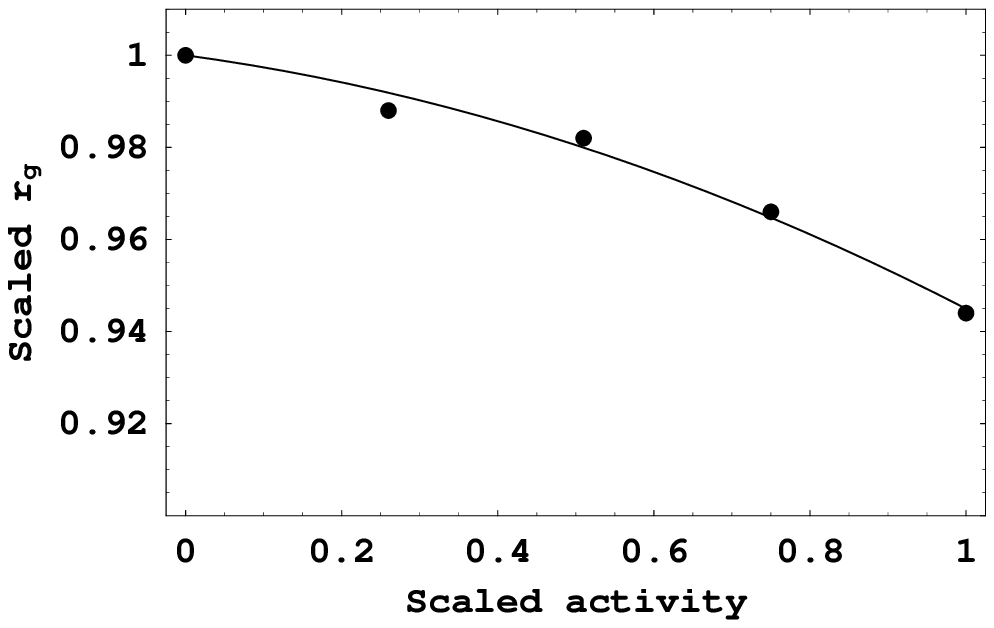}}
\par\end{centering}

\caption{\label{fig:GrowthRateReduction}Variation of the specific growth rate
due to induction of \emph{lac}. (a)~During growth of \emph{E. coli}
B in the presence of succinate and TMG, full induction of the cells
reduces the specific growth by $\sim$30\%~\citep[Table 2]{novick57}.
(b)~During growth of \emph{E. coli} K12 MG1655 in the presence of
glycerol and IPTG, full induction reduces the specific growth rate
by only $\sim$5\%~\citep[Fig.~2a]{Dekel2005}.}

\end{figure}

\paragraph*{Dependence of $r_{g,0}$ on the enzyme activity}

The model assumes that in the absence of glucose, the specific growth
rate is a fixed constant, $r_{g,0}$. However, the data shows that
the specific growth rate varies significantly with the enzyme level
of the cells (Fig.~\ref{fig:GrowthRateReduction}a). When the cells
are exposed to 0.5~mM TMG, the specific growth rate is $\sim$30\%
lower than that of non-induced cells, and the graph suggests that
the specific growth rate declines further at higher extracellular
TMG levels. The reduced specific growth rate of the induced cells
serves to decrease the off threshold because the lower the specific
growth rate, the higher the enzyme and intracellular TMG levels.

In earlier work, we studied the dynamics of growth on lactose~\citep{Narang2007c}.
There, we showed that bistability is suppressed during growth on lactose
because the specific growth rate increases with the enzyme level,
thus enhancing the stabilizing effect of dilution. Since the specific
growth rate decreases with the enzyme level during growth on TMG,
it is plausible to expect that in this case, the stabilizing effect
of dilution is depressed, and the bistable regime is enlarged.

\subsubsection{Comparison of the model simulations with the data for IPTG}

The above arguments suggest that the discrepancy between the data
and the simulations would be reduced if the experiment was performed
in the absence of glucose with a gratuitous inducer that does not
result in a significant reduction of the specific growth rate. This
conclusion seems to be consistent with the data.

\begin{figure}[t]
\noindent \begin{centering}
\subfigure[]{\includegraphics[width=2.5in]{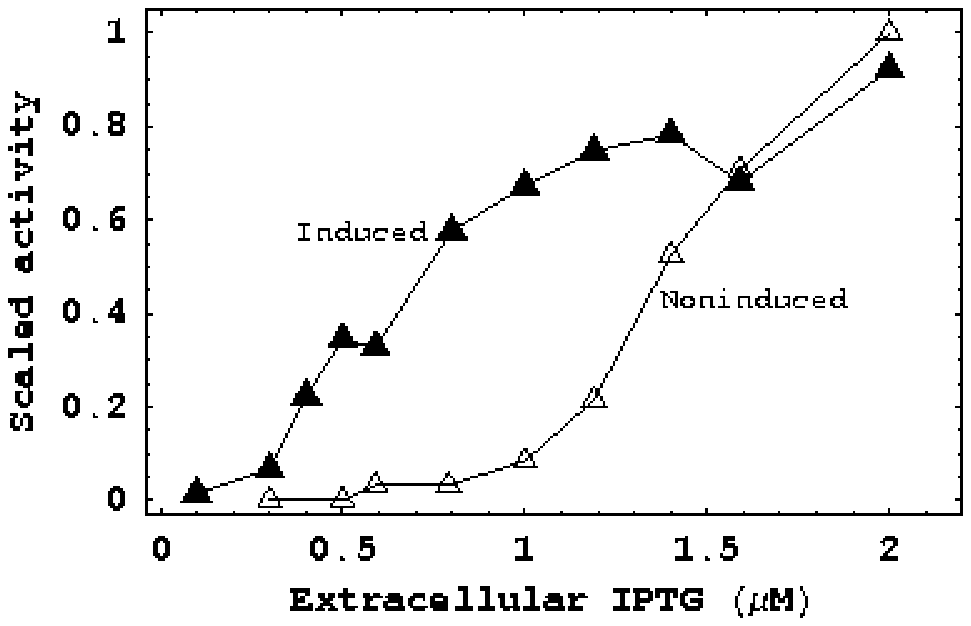}}\hspace*{0.2in}\subfigure[]{\includegraphics[width=2.5in]{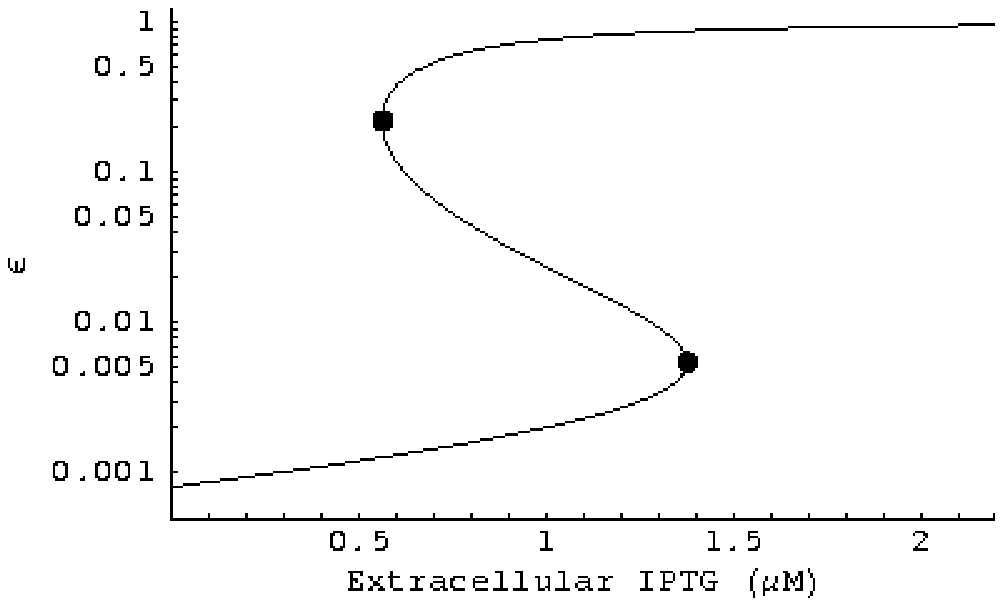}}
\par\end{centering}

\caption{\label{fig:LaurentData}Comparison of the data and the simulation
for bistability with IPTG. (a)~Variation of the observed steady state
$\beta$-galactosidase activity of \emph{E. coli} K12 with extracellular
IPTG levels at $G=0$~\citep[Fig.~7]{Laurent2005}. (b)~Variation
of the simulated enzyme activity with extracellular IPTG levels at
$G=0$. The dissociation constant, $K_{x}^{-1}$, was assumed to be
7~$\mu$M. All other parameters were assumed to be the same as those
used for simulating the experiments with TMG.}

\end{figure}

It turns out that when the cells are grown in the presence of IPTG,
the specific growth rate of the induced cells is only $\sim$5\% smaller
than the specific growth rate of the non-induced cells (Fig.~\ref{fig:GrowthRateReduction}b).
Recently, Laurent et al.~measured the variation of the enzyme activity
with the extracellular IPTG levels in the absence of glucose (Fig.~\ref{fig:LaurentData}a).
They found that the mean activity does not display the discrete all-or-none
response observed by Ozbudak et al.~(presumably due to the physiological
heterogeneity of the cells and the inherent stochasticity of the induction
process). However, there is clear evidence of bistability in the 5-fold
range, 0.3--1.5~$\mu$M, which is significantly smaller than the
10-fold range shown in Fig.~\ref{f:Oudenaarden1}a. The IPTG concentrations
supporting bistability are significantly smaller than those observed
by Ozbudak et al --- the off and on thresholds are 0.3 and 1.5~$\mu$M
(compared to 3 and 30~$\mu$M in Fig.~\ref{f:Oudenaarden1}a). This
is because the affinity of IPTG for the repressor is an order of magnitude
higher than that for TMG. Simulation of the model with $K_{x}^{-1}=7$~$\mu$M
yields results that are in reasonable agreement with the data obtained
by Laurent et al (Fig.~\ref{fig:LaurentData}b).

It remains to be seen if an extension of the model, accounting for
the dependence of $\phi_{g}$ and $r_{g,0}$ on the prevailing enzyme
level, can resolve the discrepancy with respect to the data for TMG.

\section{Conclusions}

Analysis of the experimental data shows that the diffusive influx
and carrier efflux have a profound effect on inducer accumulation
in non-induced and induced cells, respectively. Since bistability
entails the coexistence of steady states corresponding to both non-induced
and induced cells, neither one of these fluxes can be neglected. We
analyzed a model of \emph{lac} bistability taking due account of the
diffusive influx and carrier efflux. We find that:

\begin{enumerate}
\item The diffusive influx decreases the on threshold.
\item The carrier influx increases the off threshold.
\item Over a wide range of permease activities, the diffusive influx has
no effect on the off threshold, and the carrier efflux has no effect
on the on threshold. Since each of these fluxes influences only one
of the thresholds, the combined effect of these fluxes can be captured
by simple analytical expressions corresponding to two limiting cases.
\item Simulations of the model show good agreement with the data for IPTG.
\item There is a significant discrepancy with respect to the data for TMG.
Analysis of the data suggests that this discrepancy is due to the
dependence of the inducer exclusion effect and the specific growth
rate on the prevailing lactose enzyme levels.
\end{enumerate}
\begin{ack}
This research was supported in part with funds from the National Science
Foundation under contract NSF DMS-0517954. We are grateful to Stefan
Oehler (IMBB-FoRTH) for discussions regarding the regulation of the
\emph{lac} operon.
\end{ack}
\bibliographystyle{C:/texmf/bibtex/bst/elsevier/elsart-harv}

\appendix

\section{\label{app:BifnCondn}Necessary condition for a bifurcation}

The Jacobian of Eqs.~\eqref{eq:x}--\eqref{eq:r} at any steady state
is\[
\left[\begin{array}{cccc}
-\frac{1}{\tau_{x}}\left(\frac{\phi\delta_{m0}f}{\kappa_{2}}+1\right)-\phi_{g} & \frac{\phi_{s}\delta_{m0}}{\tau_{x}}\left(\frac{\sigma}{\sigma+\kappa_{1}}-\frac{\chi}{\kappa_{2}}\right) & 0 & 0\\
\phi_{e}f_{\chi} & -\phi_{g} & 0 & 0\\
\phi_{e}h_{\chi} & 0 & -\phi_{g} & 0\\
0 & 0 & 0 & -\phi_{g}\end{array}\right].\]
Evidently, two of the eigenvalues are $-\phi_{g}$. Hence, the steady
state is stable if and only if the remaining two eigenvalues corresponding
to the submatrix \[
\left[\begin{array}{cc}
-\frac{1}{\tau_{x}}\left(\frac{\phi\delta_{m0}f}{\kappa_{2}}+1\right)-\phi_{g} & \frac{\phi_{s}\delta_{m0}}{\tau_{x}}\left(\frac{\sigma}{\sigma+\kappa_{1}}-\frac{\chi}{\kappa_{2}}\right)\\
\phi_{e}f_{\chi} & -\phi_{g}\end{array}\right]\]
have negative real parts. Since the trace is always negative, stability
results whenever the determinant is positive. Hence, the necessary
condition for a bifurcation is that the determinant be zero, that
is, \[
\frac{\phi_{g}}{\tau_{x}}\left(\frac{\phi\delta_{m0}f}{\kappa_{2}}+1\right)+\phi_{g}^{2}=\frac{\phi_{s}\phi_{e}f_{\chi}\delta_{m0}}{\tau_{x}}\left(\frac{\sigma}{\sigma+\kappa_{1}}-\frac{\chi}{\kappa_{2}}\right).\]
Since $\phi_{g}^{2}\ll\phi_{g}/\tau_{x}$, the bifurcation condition
can be approximated by the relation \[
1+\frac{\delta_{m}f}{\kappa_{2}}-f_{\chi}\delta_{m}\left(\frac{\sigma}{\sigma+\kappa_{1}}-\frac{\chi}{\kappa_{2}}\right)=1+\delta_{m}\left[\frac{1}{\kappa_{2}}(f+\chi f_{\chi})-\frac{\sigma}{\sigma+\kappa_{1}}f_{\chi}\right]=0,\]
where $\phi\equiv\phi_{s}\phi_{e}/\phi_{g}$ and $\delta_{m}\equiv\phi\delta_{m0}$.
This is the same as Eq.~\eqref{eq:BifnCondn}.

The asymptotic dynamics of Eqs.~\eqref{eq:x}--\eqref{eq:r} reduces
to the stability analysis of its steady states. Indeed, Eqs.~\eqref{eq:x}--\eqref{eq:e}
are uncoupled from equations \eqref{eq:g}--\eqref{eq:r}. A direct
calculation shows that the subsystem \eqref{eq:x}--\eqref{eq:e}
has a negative divergence (i.e. the trace of the Jacobian is negative)
in the entire state space. An application of the Bendixson-Dulac criterion
eliminates the possibility of periodic orbits and saddle connections,
hence any solution $(\chi,\epsilon)$ of \eqref{eq:x}--\eqref{eq:e}
converges to some steady state. It follows immediately that the solution
$(\gamma,\rho)$ of \eqref{eq:g}--\eqref{eq:r} also converges to
the corresponding steady state.

\section{\label{app:AnalysisCase1}The positive roots of Eq.~\eqref{eq:Case1ChiEqn}}

\begin{figure}[t]
\noindent \begin{centering}
\includegraphics[width=3in]{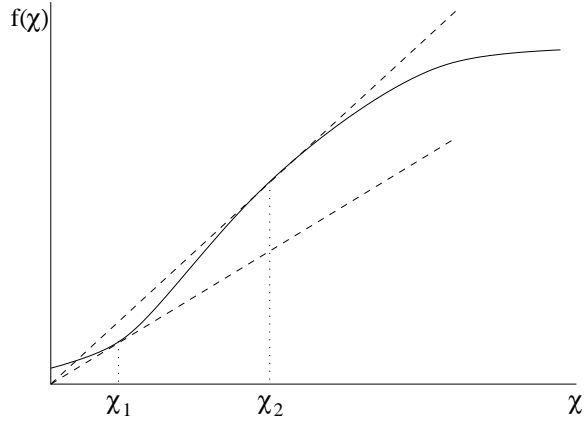}
\par\end{centering}

\caption{\label{fig:GeometryCase1}There are exactly two lines passing through
the origin (dashed lines) that are tangential to the graph of $f$
(full curve). The slopes of these lines are $f(\chi_{1})/\chi_{1}=f_{\chi}(\chi_{1})$
and $f(\chi_{2})/\chi_{2}=f_{\chi}(\chi_{2})$. }

\end{figure}

We wish to show that $q(\chi)\equiv\chi-f/f_{\chi}$ has exactly 2
positive roots, $\chi_{1}<\chi_{2}$, and is positive if and only
if $\chi_{1}<\chi<\chi_{2}$. This assertion has the a simple geometric
interpretation. Since $f(\chi)$ is a sigmoidal function for sufficiently
large $\alpha$ and $\hat{\alpha}$, there are only two lines passing
through the origin that touch the graph of $f$, and they do so at
the two points, $\chi_{1}$ and $\chi_{2}$ (Fig.~\ref{fig:GeometryCase1}).
Furthermore, for any $\chi_{1}<\chi<\chi_{2}$, the slope of the line
passing through $(0,0)$ and $(\chi,f(\chi))$ is smaller than $f_{\chi}$.

The formal proof is obtained by observing that $q(0)=-f(0)/f_{\chi}(0)<0$,
and at large $\chi$,\[
q(\chi)=\chi-\frac{1+\chi}{2}\frac{1+\alpha/\left(1+\chi\right)^{2}+\hat{\alpha}/\left(1+\chi\right)^{4}}{\alpha/\left(1+\chi\right)^{2}+2\hat{\alpha}/\left(1+\chi\right)^{4}}\approx\chi-\frac{\chi}{2}\frac{1}{\alpha/\chi^{2}}<0.\]
It follows that $q(\chi)$ has at least 2 positive roots. In fact,
it has exactly 2 positive roots because it has only 1 extremum on
$[0,\infty)$. Indeed,\[
\frac{dq}{d\chi}=1-\frac{f_{\chi\chi}f-f_{\chi}^{2}}{f_{\chi}^{2}}=\frac{f_{\chi\chi}f}{f_{\chi}^{2}},\]
which is zero if and only if $f_{\chi\chi}=0$, i.e., $\chi=\chi_{*}$,
where $\left(\chi_{*},f(\chi_{*})\right)$ is the unique inflection
point of the sigmoidal function, $f$.

\section{\label{app:AnalysisCase2}Bifurcation diagram for diffusive influx}

\begin{figure}[t]
\noindent \begin{centering}
\includegraphics[width=3in]{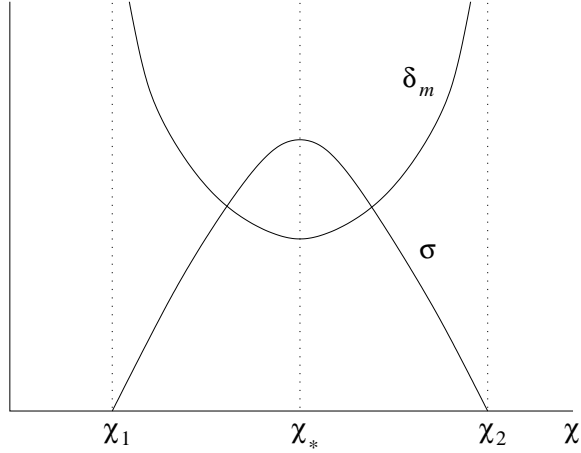}
\par\end{centering}

\caption{\label{fig:GeometryCase2}Geometry of the $\delta_{m}(\chi)$ and
$\sigma(\chi)$ in the case of diffusive influx.}

\end{figure}

We wish to show that regardless of the parameter values, the functions,
$\sigma(\chi)$ and $\delta_{m}(\chi)$, defined by \eqref{eq:Case2_Sigma}
and \eqref{eq:Case2_DeltaM}, respectively, have the geometry shown
in Fig.~\ref{fig:GeometryCase2}. It follows that as $\chi$ increases
from $\chi_{1}$, $\sigma$ increases and $\delta_{m}$ decreases
until a cusp forms at the point $\chi=\chi_{*}$ where $d\sigma/d\chi=d\delta_{m}/d\chi=0$.
Beyond $\chi=\chi_{*}$, $\sigma$ decreases and $\delta_{m}$ increases
until they approach 0 and $\infty$, respectively. Thus, the variation
of the thresholds on the $\delta_{m}\sigma$-plane has the geometry
shown in Fig.~\ref{fig:Case2BD}a, regardless of the parameter values.

Now, the geometry of $\sigma(\chi)$ is identical to that of $q(\chi)$,
which was analyzed in Appendix~\ref{app:AnalysisCase1}. Hence, it
suffices to show that $\delta_{m}(\chi)$ has the geometry shown in
Fig.~\ref{fig:GeometryCase2}. To this end, observe that \eqref{eq:Case2_DeltaM}
implies $\lim_{\chi\rightarrow\chi_{1}^{+},\chi_{2}^{-}}\delta_{m}(\chi)=\infty$.
Moreover, since\begin{align*}
\frac{d\delta_{m}}{d\chi} & =-\frac{\kappa_{1}}{\sigma^{2}}\frac{d\sigma}{d\chi}-\left(1+\frac{\kappa_{1}}{\sigma}\right)\frac{f_{\chi\chi}}{f_{\chi}^{2}},\\
 & =-\frac{d\sigma}{d\chi}\left[\frac{\kappa_{1}}{\sigma^{2}f_{\chi}}+\left(1+\frac{\kappa_{1}}{\sigma}\right)\frac{1}{f}\right],\end{align*}
$d\delta_{m}/d\chi$ and $d\sigma/d\chi$ have opposite signs at all
$\chi_{1}<\chi<\chi_{2}$, except when $\chi=\chi_{*}$, at which
point $d\delta_{m}/d\chi=d\sigma/d\chi=0$.

\section{\label{app:AnalysisCase3}Bifurcation diagram for carrier influx}

\begin{figure}[t]
\noindent \begin{centering}
\includegraphics[width=3in]{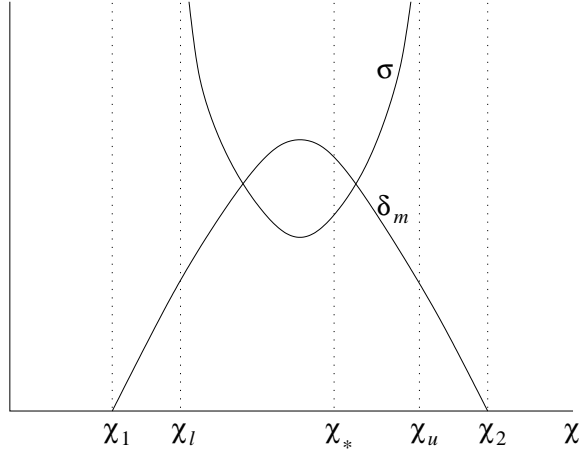}
\par\end{centering}

\caption{\label{fig:GeometryCase3}Geometry of the $\delta_{m}(\chi)$ and
$\sigma(\chi)$ in the case of carrier efflux.}

\end{figure}

We show below that regardless of the parameter values, $\delta_{m}(\chi)$
and $\sigma(\chi)$, as defined by \eqref{eq:Case3_DeltaM} and \eqref{eq:Case3_Sigma}
have the geometry shown in Fig.~\ref{fig:GeometryCase3}. Hence,
the variation of the thresholds on the $\delta_{m}\sigma$-plane always
has the geometry shown in Fig.~\ref{fig:Case3BD}a.

To see this, observe that since $q(\chi)\equiv\chi-f_{\chi}/f$ has
exactly two positive zeros, $\chi_{1}<\chi_{2}$ (Appendix~\ref{app:AnalysisCase1}),
so does $\delta_{m}(\chi)$. Furthermore, $\delta_{m}(\chi)$ has
a unique maximum on $\chi_{1}\le\chi\le\chi_{2}$. Indeed, if we let
$z\equiv\chi+1$, $\delta_{m}(\chi)$ can be rewritten as \[
\delta_{m}(z)=-\kappa_{2}\left[(z-1)\left(\frac{1}{f}\right)_{z}+\frac{1}{f}\right]=-\kappa_{2}\left[\frac{2\alpha}{z^{3}}+\frac{4\hat{\alpha}}{z^{5}}+1-\frac{\alpha}{z^{2}}-\frac{3\hat{\alpha}}{z^{4}}\right],\]
which implies that \[
\frac{d\delta_{m}(z)}{dz}=\frac{2\kappa_{2}}{z^{6}}\left[3\alpha z^{2}+10\hat{\alpha}-\alpha z^{3}-6\hat{\alpha}z\right]\equiv\frac{2\kappa_{2}}{z^{6}}b(z).\]
Now, the polynomial $b(z)$ changes sign exactly once in the interval
$(1,+\infty)$ because $b(1)=4\alpha+8\hat{\alpha}>0$, $b(z)<0$
for large $z$, and $b_{zz}(z)=6\alpha(1-z)<0$ for all $z>1$ (i.e.
$b(z)$ is a concave function). Thus, the function $\delta_{m}(\chi)$
has a unique maximum.

The function, $\sigma(\chi)$, becomes infinite at points, $0<\chi_{l}<\chi_{u}$,
lying within $(\chi_{1},\chi_{2})$. Indeed, \eqref{eq:Case3_Sigma}
yields\[
\frac{\kappa_{1}}{\sigma}=\frac{f^{2}\delta_{m}(\chi)}{\chi^{2}f_{\chi}}-1\]
so that $\sigma\rightarrow\infty$ if and only if\[
\frac{\chi^{2}f_{\chi}}{f^{2}}=\kappa_{2}\left(\frac{\chi f_{\chi}}{f^{2}}-\frac{1}{f}\right)\Leftrightarrow\frac{\chi f_{\chi}}{f}\left(1-\frac{\chi}{\kappa_{2}}\right)=1.\]
Now, $\delta_{m}(\chi)$ becomes zero if and only if $\chi f_{\chi}/f=1$.
It follows that $\kappa_{2}$ is sufficiently large, $\sigma\rightarrow\infty$
at two points, $\chi_{l}<\chi_{u}$ lying within $\left(\chi_{1},\chi_{2}\right)$.

Finally, $\sigma(\chi)$ has exactly 1 extremum in $\left(\chi_{l},\chi_{u}\right)$
which is attained at the very same value of $\chi$ that maximizes
$\delta_{m}$. Indeed,\[
\delta_{m}\frac{\kappa_{1}}{\left(\kappa_{1}+\sigma\right)^{2}}\frac{d\sigma}{d\chi}=-\left(\frac{\sigma}{\kappa_{1}+\sigma}-\frac{\chi}{\kappa_{2}}\right)\frac{d\delta_{m}}{d\chi},\]
where the term in parentheses is positive whenever $\delta_{m}(\chi)>0$
because\[
\frac{\sigma(\chi)}{\kappa_{1}+\sigma(\chi)}=\frac{\chi^{2}f}{\kappa_{2}(\chi f_{\chi}-f)}\Rightarrow\frac{\sigma(\chi)}{\kappa_{1}+\sigma(\chi)}-\frac{\chi}{\kappa_{2}}=\frac{\chi f}{\kappa_{2}(\chi f_{\chi}-f)}.\]
It follows that $d\delta_{m}/d\chi$ and $d\sigma/d\chi$ have opposite
signs in the interval $\chi_{1}<\chi<\chi_{2}$, and the functions
$\delta_{m}$ and $\sigma$ attain their extrema at the same values
of $\chi$. Since $\delta_{m}$ attains a unique maximum in the interval
$\chi_{1}<\chi<\chi_{2}$, so does $\sigma$.
\end{document}